\documentclass[twocolumn,aps,prb,longbibliography,superscriptaddress,floatfix]{revtex4-2}

\usepackage[T1]{fontenc}
\usepackage[utf8]{inputenc}
\usepackage[english]{babel}
\usepackage{color}
\usepackage{float}
\usepackage{braket}
\usepackage{amsmath}
\usepackage{amstext}
\usepackage{amssymb}
\usepackage{graphicx}
\usepackage{bbold}
\usepackage[unicode=true,pdfusetitle,
bookmarks=true,bookmarksnumbered=false,bookmarksopen=false,
breaklinks=false,pdfborder={0 0 1},backref=false,colorlinks=false]
{hyperref}
\newcommand{\angstrom}{\textup{\AA}}
\hypersetup{
	colorlinks,linkcolor=blue,citecolor=blue,urlcolor=blue}
\usepackage{soul}
\makeatletter
\date{}

\makeatother
\setul{0.5ex}{0.3ex}
\definecolor{Blue}{rgb}{0,0.0,1}
\setulcolor{Blue}
\usepackage{babel}

\begin{document} 

\author{Tarik P. Cysne}
\email{tarik.cysne@gmail.com}
\affiliation{Instituto de F\'\i sica, Universidade Federal Fluminense, 24210-346 Niter\'oi RJ, Brazil}

\author{W. J. M. Kort-Kamp}
\email{kortkamp@lanl.gov}
\affiliation{Theoretical Division, Los Alamos National Laboratory, Los Alamos, New Mexico 87545, USA}

\author{Tatiana G. Rappoport}
\email{tgrappoport@gmail.com}
\affiliation{Physics Center of Minho and Porto Universities (CF-UM-UP), Braga, Portugal}	
\affiliation{Instituto de F\'\i sica, Universidade Federal do Rio de Janeiro, C.P. 68528, 21941-972 Rio de Janeiro RJ, Brazil} 

\title{Controlling the orbital Hall effect in gapped bilayer graphene in the terahertz regime}

\begin{abstract}
We study the orbital Hall effect (OHE) in the AC regime using bilayer graphene (BLG) as a prototypical material platform. While the unbiased BLG has gapless electronic spectra, applying a perpendicular electric field creates an energy band gap that can be continuously tuned from zero to high values. By exploiting this flexibility, we demonstrate the ability to control the behavior of AC orbital Hall conductivity. Particularly, we demonstrate that the orbital Hall conductivity at the neutrality point changes its signal at a critical frequency, the value of which is proportional to the perpendicular electric field. For BLG with narrow band gaps, the active frequency region for the AC OHE may extend to a few terahertz, which is experimentally accessible with current technologies. We also consider the introduction of a perpendicular magnetic field in the weak coupling regime using first-order perturbation theory to illustrate how the breaking of time-reversal symmetry enables the emergence of AC charge Hall effect in the charge-doped situation and modifies the AC orbital Hall conductivity. Our calculations suggest that BLG with narrow bandgaps is a practical candidate for investigating time-dependent orbital angular momentum transport.
\end{abstract} 
\maketitle


\section{Introduction} 

In recent years, there has been a growing interest in the electronic transport of orbital angular momentum (OAM) in materials. Orbital currents can be generated in low spin-orbit coupled materials through electric means, thanks to the phenomenon of the orbital Hall effect (OHE). This effect was predicted almost twenty years ago \cite{Bernevig-PhysRevLett.95.066601} but only recently it was directly probed in Ti and other light metals \cite{Nature-OrbitalHall-Hyun-Woo, DetectionOHE-PhysRevLett.131.156702, DetectionOHE-PhysRevLett.131.156703}. This has unlocked the potential of utilizing the OAM of electrons for information processing in devices and has given rise to the emerging field of orbitronics \cite{Go-EPL-MiniReview_2021, urazhdin2023symmetry, Go-Textures-PhysRevLett.121.086602, Gambardella-PhysRevResearch.4.033037, atencia2023nonconservation, ArnabBose-PhysRevB.107.134423, PhysRevB.105.235142, Oppeneer-PhysRevMaterials.6.095001, Kontani-PhysRevLett.102.016601, manchon2023orbitalDiffusion, Photoinduced-OrbitalHall-Mu2021, santos2023exploring, HWLee-PhysRevLett.128.176601, Orbital-PhysRevB.107.214109, PhysRevResearch.5.043294, Hayami-PhysRevB.98.165110, Peters-PhysRevB.107.094106, go2023orbital-Pumping, Zeer-PhysRevResearch.6.013095, Oppeneer-PhysRevB.106.024410}.   

The physics of OAM has also drawn the attention of the two-dimensional (2D) materials community \cite{Sayantika-PhysRevB.102.035409, Sayantika-PhysRevB.101.121112, Us-NRpband-PhysRevB.104.165403, Us-NR-TMD-PhysRevB.107.115402, Go-Mokrousov-PhysRevMaterials.6.074004, Orbital-Chern-PhysRevB.108.224422, ji2023reversal, VHE_X_OHE-PhysRevB.102.161103, Hayami_JPCMSymmetries_2016, AndersonBarbosa-PhysRevB.108.245105, OME-PhysRevB.103.224426, ghosh2023orbital, Hayami_PhysRevB.90.081115, liu2023dominance-Extrinsic, Manchon-PhysRevB.108.075427, ARPES-PhysRevX.12.011019, ARPES-PhysRevLett.125.216404, OAMText-Schler2023}. Some of these materials exhibit non-trivial orbital textures that are responsible for the OHE at the intra-atomic level \cite{Go-Textures-PhysRevLett.121.086602, ARPES-PhysRevX.12.011019, ARPES-PhysRevLett.125.216404, OAMText-Schler2023}, and they were also predicted to host the orbital Hall insulating phase \cite{Canonico-PhysRevB.101.161409, Canonico-PhysRevB.101.075429, Us-BP-PhysRevB.108.165415}. These insulating phases exhibit intriguing topological characteristics. In certain systems, it can be indexed by the \emph{orbital Chern number} topological invariant \cite{Cysne2021-PhysRevLett.126.056601, Cysne-Bhowal-Vignale-Rappoport-PhysRevB.105.195421} and has also been recently associated with high-order topology  \cite{HOTI-TMD-PhysRevB.105.045417, Us-HO-PRL-PhysRevLett.130.116204}. One of the appealing aspects of 2D materials is the ability to control their electronic band structure through external perturbations. In this context, bilayer graphene (BLG) stands out from the other systems due to its seamless synthesis and integration in a variety of substrates and the easy tunability of its band structure \cite{Gao2020-ReviewBG, Yan2015-ReviewBG_Photonics}. Indeed, the unbiased BLG exhibits gapless electronic spectra, with conduction and valence bands touching each other at the Dirac points. The application of an electric field perpendicular to the BLG opens a band gap at these points, which can be continuously tuned from zero up to $\approx 250$ meV \cite{Gao2020-ReviewBG}. This alters the spectral dispersion of electrons, making it a practical material platform for exploring the properties associated with the quantum geometry of Bloch states across the Brillouin zone. The geometrical properties of electrons encoded in Berry curvature, which strongly depend upon the energy gaps, have indeed been investigated to attain a high photoinduced Hall effect in BLG with narrow bandgaps \cite{Koppens-Sience-2023, GiantHallPhotoconductivity-JSong_Nanolett}. The electrons in BLG exhibit $p_z$ orbital character and, consequently, do not possess intra-atomic OAM. However, a finite OAM contribution can arise from the intersite movement of electrons \cite{ISouza-modTheo-PhysRevB.77.054438, FXuan-modTheo-PhysRevResearch.2.033256}, leading to the OHE \cite{Bhowal-Vignale-PhysRevB.103.195309}. The primary contribution to intersite OAM and the OHE in BLG comes from Bloch states near the valleys of the Brillouin zone. Then, the BLG system may offer an avenue to shed light on the recent debate concerning the mechanism of orbital transport in Dirac systems \cite{Review-VHE-Roche2022, Bhowal-Vignale-PhysRevB.103.195309}, which is also characterized by its dependence on bandgap.   

The response of materials at non-zero frequencies (AC regime) can offer a deeper understanding of the mechanisms involved in electronic transport. This approach has been employed as a method to explore spintronic phenomena in both theoretical and experimental research \cite{AC-SpincurrentDetection, AC-SHE-PhysRevLett.113.157204}. Numerous calculations on the spin Hall effect at finite frequencies have been reported \cite{AC_SHE-PhysRevB.87.235426, Chen-JPCM_ACSHE_2023, AC-SHC-PhysRevLett.94.226601, Bechara-THz-2017, AC_SHE-TMD_PhysRevB.103.L161410}, thereby exploring the potential applications of spintronics in ultrafast regimes. Additionally, the advancement of laser physics has the potential to pave the way for innovative techniques to investigate the responses of materials in broad frequency ranges. It has inspired recent research in time-dependent spin and orbital dynamics \cite{MeleOrbitronics-PhysRevLett.123.236403, Mertig-PhysRevB.108.184401, Mertig-PhysRevB.108.104408, hamamera2023ultrafast}. For example, it has been demonstrated that the picture of the steady OHE can be extended to the ultrafast regime, accompanied by a reduction in the coherence of the physical quantities involved when reaching the femtosecond scale \cite{Ultrafast-OHE-Mertig}. On the picosecond timescale, achieved through the use of terahertz radiation, the approach based on conductivity tensors accurately describes transport properties in the frequency domain. The relevance of understanding the OHE at non-zero frequencies increases when considering recent experiments that demonstrate the generation of terahertz emission signals related to electronic OAM transport \cite{Fert-xu2023orbitronics, Go-Thz-emission-10.1038/s41565-023-01470-8, IOHE-THz-s41535-023-00559-6, mishra2024active}.  

Here we explore the OHE in BLG in the AC regime. We demonstrate that it is possible to manipulate the behavior of orbital Hall conductivity (OHC) in the frequency domain by adjusting the intensity of the perpendicular electric field applied in the bilayer. In particular, we show that the real part of OHC undergoes a sign reversal at a critical frequency, proportional to the applied perpendicular electric field. This critical frequency can be reduced to a few THz in the case of narrow band gaps. Additionally, we consider the effects of a weak external magnetic field on electronic transport. We demonstrate that the breaking of time-reversal symmetry induces an AC charge Hall effect in the charge-doped system. It also modifies the profile of the OHC for Fermi energy near the band edges.

\begin{figure}[h!]
	\centering
	 \includegraphics[width=1.0\linewidth,clip]{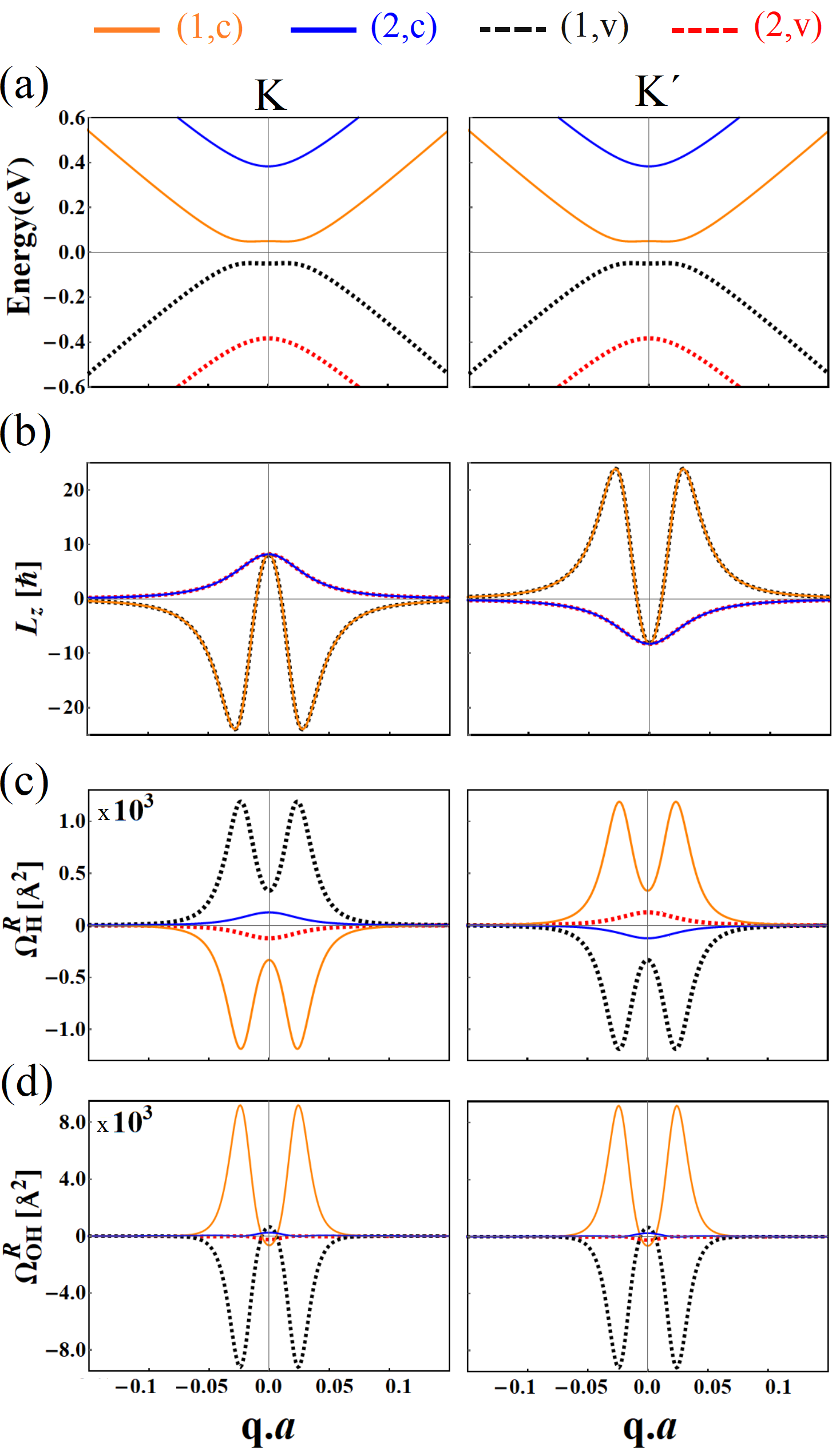}   
	\caption{(a) The electronic spectrum of BLG near the valleys [Eq. (\ref{HBGB})]. (b) The OAM of Bloch states [diagonal matrix elements of Eq.(\ref{MMDegenerateBLgraphene}) multiplied by $-\hbar/(g_L\mu_B)$] for the conduction $[(1,c); (2,c)]$ and valence $[(1,v); (2,v)]$ bands. Here, for illustrative purposes, we used a large bandgap $\Delta=0.1 \text{eV}$. (c) Berry curvature [Eq. (\ref{OmegaH}) with $B=0$ and $\hbar \omega=0$] and (d) Orbital Berry curvature [Eq. (\ref{OBC}) with $\hbar \omega=0$] of Bloch bands. The left (right) panels show the results for valley ${\bf K}$ (${\bf K'}$).}
	\label{BLGSpectra_mz}
\end{figure}

\section{Gapped bilayer graphene model \label{Sec:model}}

The intersite OAM of BLG concentrates on Bloch states near the valleys of the Brillouin zone. In the low-energy limit, we use the tight-binding (TB) basis of the Hamiltonian of BLG with Bernal stacking in the valley ${\bf K}=\hat{x}(4\pi)/(3\sqrt{3}a)$ as $\beta^{BL}_{tb}({\bf K})=\{ A_u, B_b, A_b, B_u\}$, and in the valley ${\bf K}'=-{\bf K}$ as $\beta^{BL}_{tb} ({\bf K}')=\{ B_b, A_u, B_u, A_b\}$. $A$ and $B$ refer to the sublattices of the honeycomb arrangement, and subscripts $u$ and $b$ refer to the upper and bottom layers of the BLG system. In this basis, one writes the low-energy Hamiltonian as \cite{2L-graphene-McCann-PhysRevB.74.161403}
\begin{eqnarray}
H_{\rm BL}({\bf q}_{\tau})=\begin{bmatrix}
   -\frac{\tau \Delta}{2} & 0 & 0 & \gamma_{-} \\
   0 & \frac{\tau \Delta}{2} & \gamma_{+} & 0 \\
   0 & \gamma_{-} & \frac{\tau \Delta}{2} & t_{\perp} \\
   \gamma_{+} & 0 & t_{\perp} & -\frac{\tau \Delta}{2}
\end{bmatrix},
\label{HBGB}
\end{eqnarray}
where $\gamma_{\pm}=\hbar \tilde{v}\tau(q_x\pm i q_y)$ with ${\bf q}_{\tau}={\bf k}-\tau {\bf K}$ and $\tilde{v}=3a\tilde{t}/(2\hbar)$ with $a=1.42 \angstrom$ and the renormalized nearest-neighbor hopping amplitude $\tilde{t}= 3.16 \text{eV}$. For the isolated BLG, the interlayer hopping is $t_{\perp}=0.38 \text{eV}$. $\tau$ is the quantum number associated with the valley degree of freedom and assumes values $\tau=\pm 1$ for valleys ${\bf K}$ and ${\bf K}'$, respectively. The term $\Delta$ produces an asymmetry in the on-site energy of each layer, modeling the effect of a perpendicularly applied electric field or the interaction of the BLG with a substrate.   

When $\Delta=0$, the electronic spectrum of BLG is gapless. The finite value of $\Delta$ opens a band gap at the Dirac point, as illustrated in Fig. \ref{BLGSpectra_mz} (a). It also breaks spatial inversion symmetry, endowing the Bloch bands with finite orbital magnetic moment [Fig. \ref{BLGSpectra_mz} (b)] and Berry curvature [Fig. \ref{BLGSpectra_mz} (c)]. The interplay between the valley degrees of freedom, non-trivial Berry curvature, and the orbital magnetic moment has been extensively investigated in the context of valleytronics. One may also define an orbital Berry curvature associated with the OHE [Fig. \ref{BLGSpectra_mz} (d)]. In the following, we use linear-response theory to examine the OHE in the BLG model described above under the influence of an AC driving field. We evaluate our calculations numerically, since the strong interlayer coupling in BLG and the need for flexibility in the variation of parameter $\Delta$ from zero to high values forbid perturbative expansion as done in Ref. \cite{Cysne-Bhowal-Vignale-Rappoport-PhysRevB.105.195421}.


\begin{figure}
	\centering
	 \includegraphics[width=0.75\linewidth,clip]{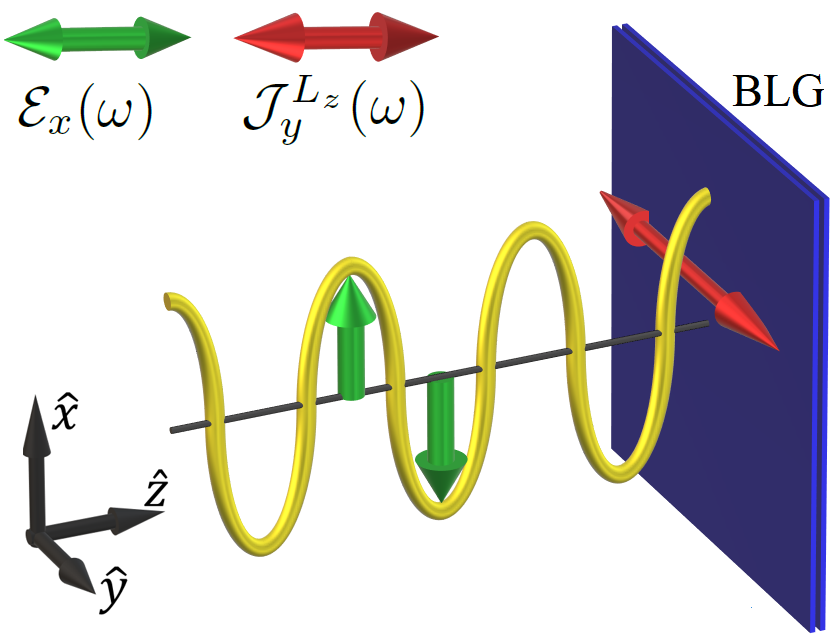}   
	\caption{Schematics of gapped BLG subject to a normal incident harmonic plane wave polarized along the x direction.}
	\label{Boneco}
\end{figure}

\section{AC orbital Hall Effect}

We consider the gapped BLG system subjected to a normally incident harmonic plane wave polarized along the x-direction, with electromagnetic field $\mathcal{E}_x(\omega)$ \cite{ISouza-modTheo-PhysRevB.77.054438}, as illustrated in Fig. \ref{Boneco}. The response quantities to this oscillating electric field can be calculated within linear response theory using the Kubo formula. The oscillating transverse orbital current generated by the AC electric field is $\mathcal{J}^{L_z}_y(\omega)=\sigma_{{\rm OH}}(\omega)\mathcal{E}_x(\omega)$, where the AC OHC $\sigma_{{\rm OH}}(\omega)=\sigma^{R}_{{\rm OH}}(\omega)+i\sigma^{I}_{{\rm OH}}(\omega)$ is given by \cite{AC_SHE-PhysRevB.87.235426, AC_SHE-TMD_PhysRevB.103.L161410}

\begin{widetext}
\begin{eqnarray}
\sigma^{R/I}_{{\rm OH}}(\omega)=e\sum_n\sum_{\tau=\pm 1}\int \frac{d^2q}{(2\pi)^2}\Omega^{R/I}_{{\rm OH},n} ({\bf q}_{\tau},\omega) \Theta(E_{\rm F}-E_{n,{\bf q}_{\tau}}), \label{sigmaOptcOH}
\end{eqnarray}
with,
\begin{eqnarray}
\frac{\Omega^{R/I}_{{\rm OH},n}({\bf q}_{\tau},\omega)}{2\hbar}=\frac{1}{2}\sum_{m\neq n}\begin{Bmatrix}
  \text{Im} \\ -\text{Re} \\\end{Bmatrix}\Bigg[\frac{1 }{(E_{n,{\bf q}_{\tau}}-E_{m,{\bf q}_{\tau}})} \Bigg(\frac{ \bra{u_{n,{\bf q}_{\tau}}}\hat{v}_x({\bf q}_{\tau})\ket{u_{m,{\bf q}_{\tau}}} \bra{u_{m,{\bf q}_{\tau}}}\hat{J}^{L_z}_y({\bf q}_{\tau})\ket{u_{n,{\bf q}_{\tau}}} }{(E_{n,{\bf q}_{\tau}}-E_{m,{\bf q}_{\tau}}+\hbar \omega+i\eta)}+(n \leftrightarrow m)\Bigg)\Bigg], \label{OBC}
\end{eqnarray} 
\end{widetext} 
where, $E_{n(m),{\bf q}_{\tau}}$ is the energy of Bloch state with periodic part $\ket{u_{n(m), {\bf q}_{\tau}}}$ and $\hat{v}_{x(y)}({\bf q}_{\tau})=\hbar^{-1}\partial \hat{H}_{\rm BL}({\bf q}_{\tau})/\partial q_{x(y)}$ is the velocity operator. $\eta$ is a small phenomenological relaxation rate that accounts for dissipative effects in the bilayer. In Eq. (\ref{sigmaOptcOH}) and throughout this paper, we employ the zero temperature limit of the Fermi-Dirac distribution given by a Heaviside function: $f_{\rm FD}(E_{\rm F}, E_{n,{\bf q}_{\tau}})\rightarrow \Theta(E_{\rm F}-E_{n,{\bf q}_{\tau}})$. In Eq. (\ref{OBC}), the OAM current operator is $\hat{J}^{L_z}_y({\bf q}_{\tau})=\left(\hat{L}_z({\bf q}_{\tau}) \hat{v}_y({\bf q}_{\tau})+\hat{v}_y({\bf q}_{\tau})\hat{L}_z({\bf q}_{\tau}) \right)/2$. Here, we follow Refs. \cite{Bhowal-Vignale-PhysRevB.103.195309, Cysne-Bhowal-Vignale-Rappoport-PhysRevB.105.195421} and describe the OAM operator as $\hat{L}_z({\bf q}_{\tau})=-(\hbar/\mu_Bg_L)\hat{\mathbb{m}}^{z,{\rm tb}}_{{\bf q}_{\tau}}$, where $g_L=1$ is the Landé g-factor and $\mu_B=e\hbar/2m_e$ is the atomic Bohr magneton in terms of electron's rest mass $m_e$. The construction of the orbital magnetic moment operator $\hat{\mathbb{m}}^{z, {\rm tb}}_{{\bf q}_{\tau}}$ in tight-biding basis is detailed in the Appendix \ref{AppendixA}. The OHC can be finite even in a system that preserves both spatial inversion and time-reversal symmetry \cite{Cysne2021-PhysRevLett.126.056601, Cysne-Bhowal-Vignale-Rappoport-PhysRevB.105.195421}. In nonzero frequencies, the AC OHC has a real ($\sigma^{R}_{{\rm OH}}(\omega)$) and an imaginary ($\sigma^{I}_{{\rm OH}}(\omega)$) part. To calculate $\sigma^{R}_{{\rm OH}}(\omega)$ ($\sigma^{I}_{{\rm OH}}(\omega)$), one takes $\text{Im}$ ($-\text{Re}$) of the quantity inside the brackets $[...]$ in Eq. \ref{OBC}.

\section{Results}

\begin{figure}
	\centering
	 \includegraphics[width=1\linewidth,clip]{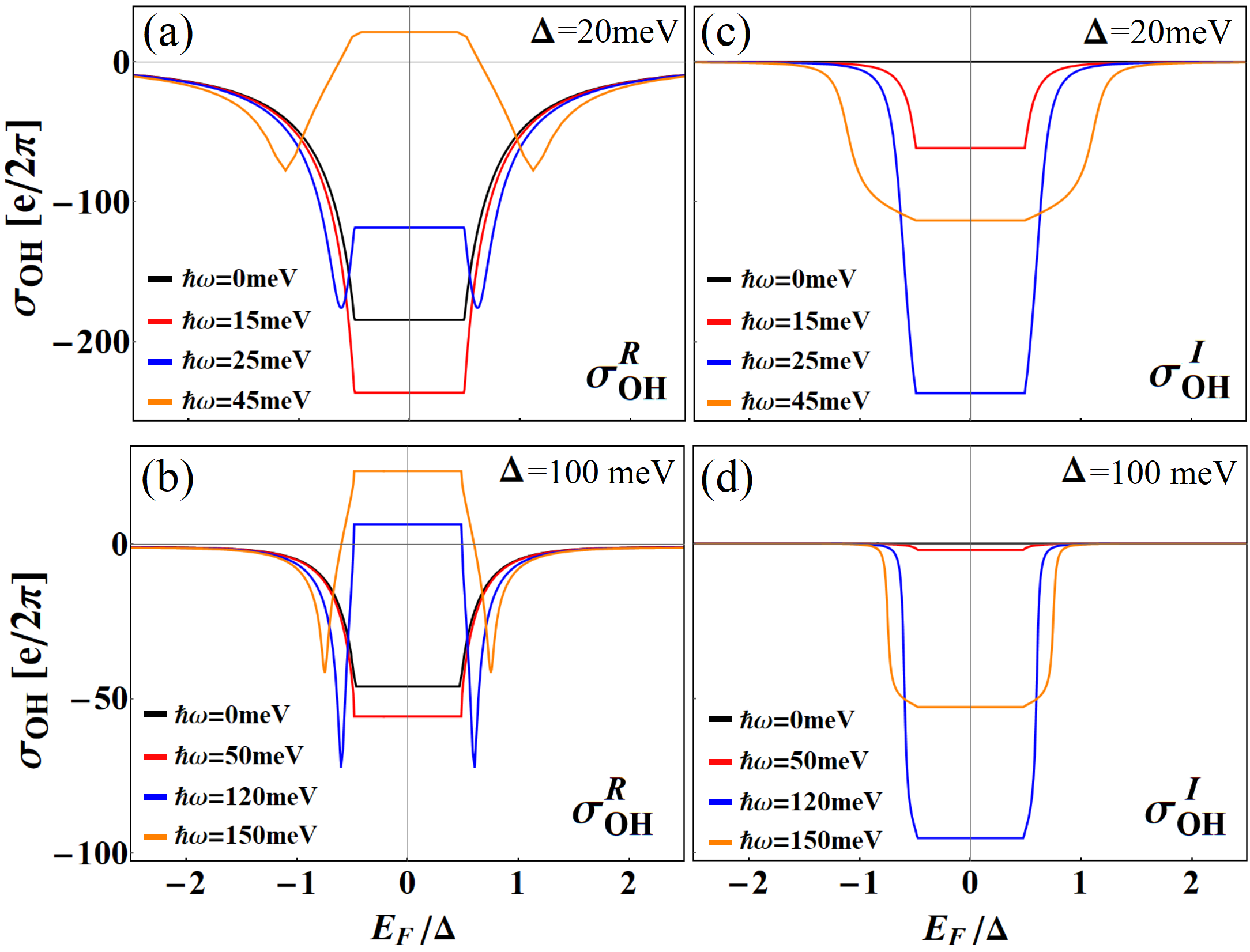}   
	\caption{(a,b) Real and (c,d) Imaginary parts of OHC as a function of $E_{\rm F}/\Delta$ for different values of frequency. We used $\Delta=20 \text{meV}$ in panels (a,c) and $\Delta=100 \text{meV}$ in panel (b,d). Here, we set $\eta=4$ meV.}
	\label{OHC_X_EF}
\end{figure} 

\begin{figure}
	\centering
	 \includegraphics[width=0.75\linewidth,clip]{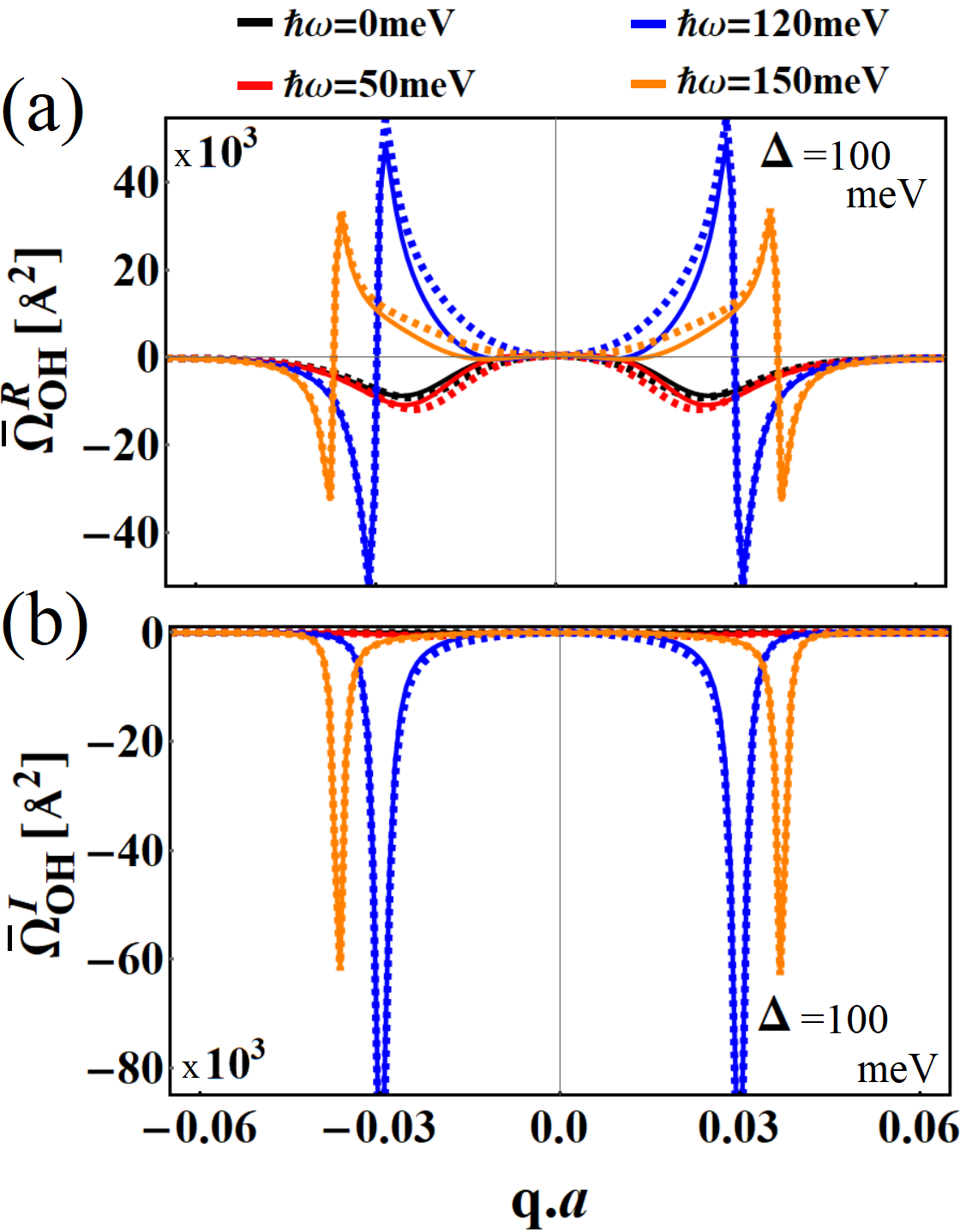}  
	\caption{Solid lines represent the (a) real and (b) imaginary parts of total orbital Berry curvatures of the valence band $\bar{\Omega}^{R/I}_{{\rm OH}}(\omega, {\bf q}_{\tau})=\sum_{n=1,2}\Omega^{R/I}_{{\rm OH}, n,{\rm v}}(\omega, {\bf q}_{\tau})$ calculated for different frequency values, for valley ${\bf K}$. Dashed lines represent the functions $f^{R/I}(\omega, {\bf q}_{\tau})=\sum_{n=1,2}\langle \hat{L}_z({\bf q}_{\tau})\rangle_{n,{\rm v}}\Omega^{R/I}_{{\rm H},n,{\rm v}}(\omega, {\bf q}_{\tau}) /(\pi \hbar)$ summed over valence-band states, for valley ${\bf K}$. The results are the same for valley ${\bf K}'$. Here we set $\Delta=100$ meV and $\eta=4$ meV.} 
	\label{OBCB} 
\end{figure}

In Fig. \ref{OHC_X_EF}, we show the real and imaginary parts of the OHC as a function of $E_{\rm F}/\Delta$ for two values of $\Delta$ and various frequencies $\hbar \omega$. In the DC regime ($\hbar \omega=0$) the imaginary part of OHC vanishes, and an orbital Hall insulating plateau in the real part of OHC is observed when the Fermi energy lies within the energy bandgap of the electronic spectra \cite{Canonico-PhysRevB.101.161409, Canonico-PhysRevB.101.075429, Cysne2021-PhysRevLett.126.056601,Us-BP-PhysRevB.108.165415,ghosh2023orbital}. The height of this DC OHC plateau increases as the band gap narrows \cite{Bhowal-Vignale-PhysRevB.103.195309, Cysne-Bhowal-Vignale-Rappoport-PhysRevB.105.195421}. Due to the absence of a Fermi surface, the OHC in the insulating phase should not be modified by the effect of dilute disorder \cite{liu2023dominance-Extrinsic, tang2024role}. 
In the AC regime ($\hbar \omega \neq 0$), the height of the plateau in the real part of OHC changes non-monotonically with the frequency of the oscillating electric field $\hbar \omega$. Initially, the height of this plateau increases with $\hbar \omega$, reaching a maximum and then starting to decrease. At high frequencies, the height of the plateau in the real part of OHC switches its sign. In the case of the imaginary part of OHC, we observe a plateau that starts from zero, reaches a maximum value, and then decreases. In contrast to the plateau in the real part of OHC, the one in the imaginary part does not change its sign.  

The origin of the change in the sign of the real part of OHC plateau with frequency $\hbar \omega$ can be attributed to the hat-shaped electronic spectra of the gapped BLG. Due to this peculiar electronic spectrum, the expected value of the OAM operator $\langle \hat{L}_z({\bf q}_{\tau})\rangle_{n,{\rm c(v)}} = \langle u_{n,{\rm c(v)}}\big| \hat{L}_z({\bf q}_{\tau})\big| u_{n,{\rm c(v)}} \rangle$ associated with the Bloch-state $\big|u_{n=1,c(v)}\rangle$ changes its sign when the modulus of crystal momenta relative to valleys $\big| {\bf q}_{\tau}\big|$ varies from zero to large values. Specifically, at valley ${\bf K}$, $\langle \hat{L}_z({\bf q}_{\tau})\rangle_{1,{\rm c(v)}}$ is positive for small $\big| {\bf q}_{\tau}\big|$ values and becomes negative for large $\big| {\bf q}_{\tau}\big|$ values after passing through a sign inversion [see Fig. \ref{BLGSpectra_mz} (b)]. The opposite occurs at valley ${\bf K}'$. The expectation value of the OAM operator influences the orbital Berry curvature. It was demonstrated in Ref. \cite{Bhowal-Vignale-PhysRevB.103.195309} that the orbital Berry curvature of monolayer graphene for $\hbar \omega=0$ is related to the expected value of the OAM operator and the electronic Berry curvature through a mathematical relation $\Omega^{1L}_{\rm OH}({\bf q}_{\tau})=L_z({\bf q}_{\tau})\Omega^{1L}_{\rm H}({\bf q}_{\tau})/(\hbar \pi)$. One can numerically verify that a similar relation is approximately fulfilled in the case of gapped BLG in the AC regime. We define the orbital Berry curvature of the valence band of gapped BLG as $\bar{\Omega}^{R/I}_{{\rm OH}}(\omega, {\bf q}_{\tau})=\sum_{n=1,2}\Omega^{R/I}_{{\rm OH}, n,{\rm v}}(\omega, {\bf q}_{\tau})$, represented by solid lines for different values of $\hbar \omega$ in Fig. \ref{OBCB}. The dashed lines in Fig. \ref{OBCB} represent the plot of the function $f^{R/I}(\omega,{\bf q}_{\tau})= \sum_{n=1,2}\langle \hat{L}_z({\bf q}_{\tau})\rangle_{n,{\rm v}}\Omega^{R/I}_{{\rm H},n,{\rm v}}(\omega, {\bf q}_{\tau}) /(\pi \hbar)$, where $\Omega^{R/I}_{{\rm H},n,{\rm v}}(\omega, {\bf q}_{\tau})$ is the AC electronic Berry curvature defined in next section (see Eq. \ref{OmegaH}). Upon inspecting the results in Fig. \ref{OBCB}, one notices that $\bar{\Omega}^{R/I}_{{\rm OH}}(\omega, {\bf q}_{\tau})\approx f^{R/I}(\omega,{\bf q}_{\tau})$, thus establishing a connection between $\bar{\Omega}^{R/I}_{{\rm OH}}$ and $\langle \hat{L}_z({\bf q}_{\tau})\rangle$. The function $\bar{\Omega}^{R}_{{\rm OH}}$ spreads across a relatively large region of crystalline momentum [Fig. \ref{OBCB} (a)]. At small frequencies $\hbar \omega$, $\bar{\Omega}^{R}_{{\rm OH}}$ is dominated by optical transitions between Bloch states at small values of $\big| {\bf q}_{\tau}\big|$, for which $\langle \hat{L}_z({\bf q}_{\tau})\rangle > 0$ at ${\bf K}$ and $\langle \hat{L}_z({\bf q}_{\tau})\rangle < 0$ at ${\bf K}'$. At high frequencies $\hbar \omega$, $\bar{\Omega}^{R}_{{\rm OH}}$ is dominated by optical transitions between Bloch states at high values of $\big| {\bf q}_{\tau}\big|$, for which $\langle \hat{L}_z({\bf q}_{\tau})\rangle < 0$ at ${\bf K}$ and $\langle \hat{L}_z({\bf q}_{\tau})\rangle > 0$ at ${\bf K}'$. This leads to a change in the signs of the $\bar{\Omega}^{R}_{{\rm OH}}$ and in the real part of OHC plateau as $\hbar \omega$ increases. The imaginary part of the AC orbital Berry curvature $\bar{\Omega}^{I}_{{\rm OH}}$ [Fig. \ref{OBCB} (b)] is a negative peaked function around a given crystal momentum $\pm q^*(\omega)$ and maintains its sign throughout variations in frequency. The imaginary part of conductivity $\sigma^{I}_{\rm OH}(\omega)$ does not change its sign with frequency but is related to the real $\sigma^{R}_{\rm OH}(\omega)$ part by causality constraints (Kramers-Kronig relations).

\begin{figure}
	\centering
	 \includegraphics[width=0.85\linewidth,clip]{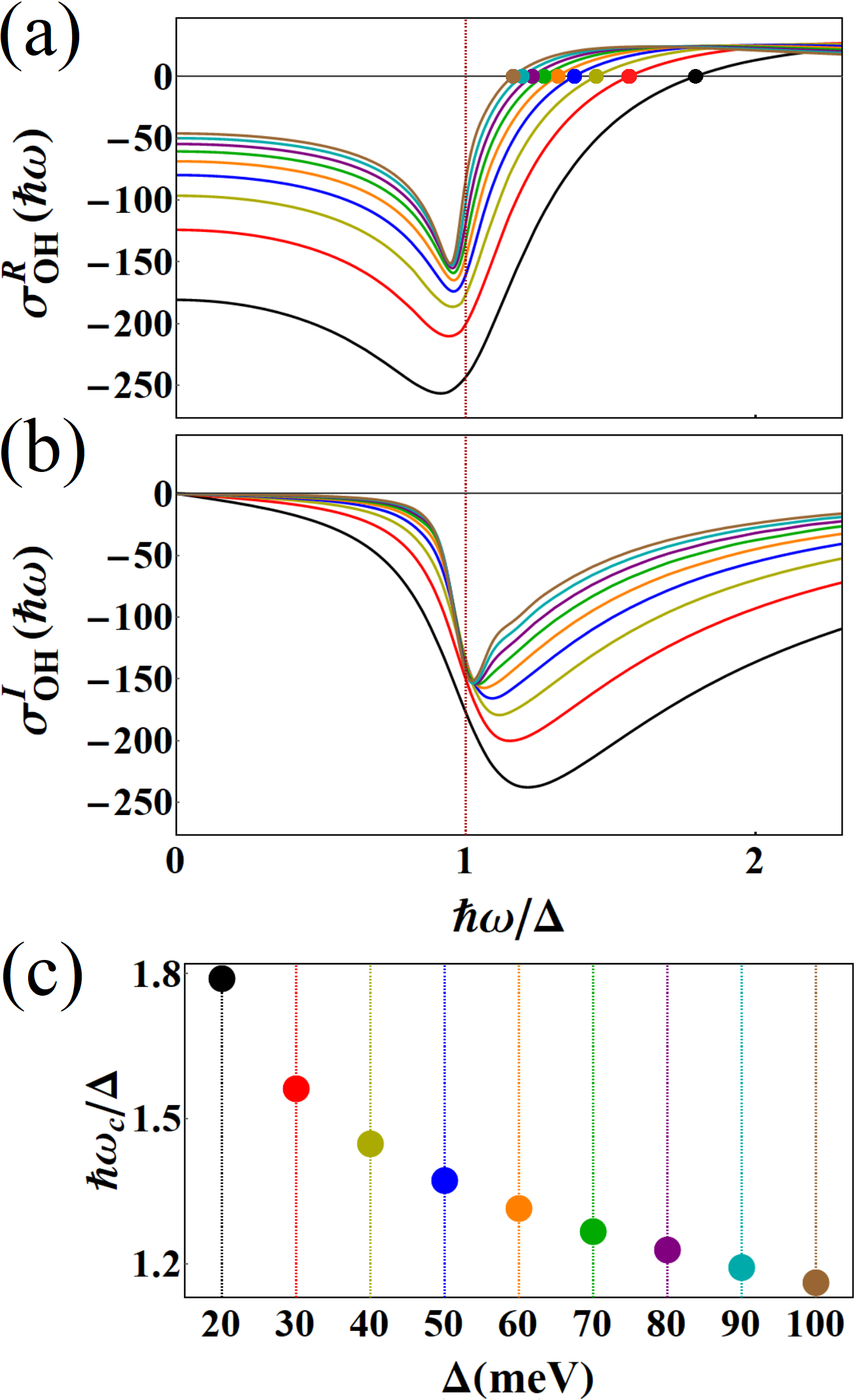}   
	\caption{(a) Real and (b) Imaginary parts of AC OHC as a function of $\hbar\omega/\Delta$ at $E_{\rm F}=0$ for different values of $\Delta$. (c) The critical value of $\hbar\omega_c/\Delta$ [circles in panel (a)] in which the real part of OHC changes its sign as a function of $\Delta$. The color code for different values of $\Delta$ used in the figure can be identified in panel (c). Here, we used $\eta=4$ meV.}
	\label{Condutividade_X_HBarOmega}
\end{figure}

\begin{figure}
	\centering
	 \includegraphics[width=0.85\linewidth,clip]{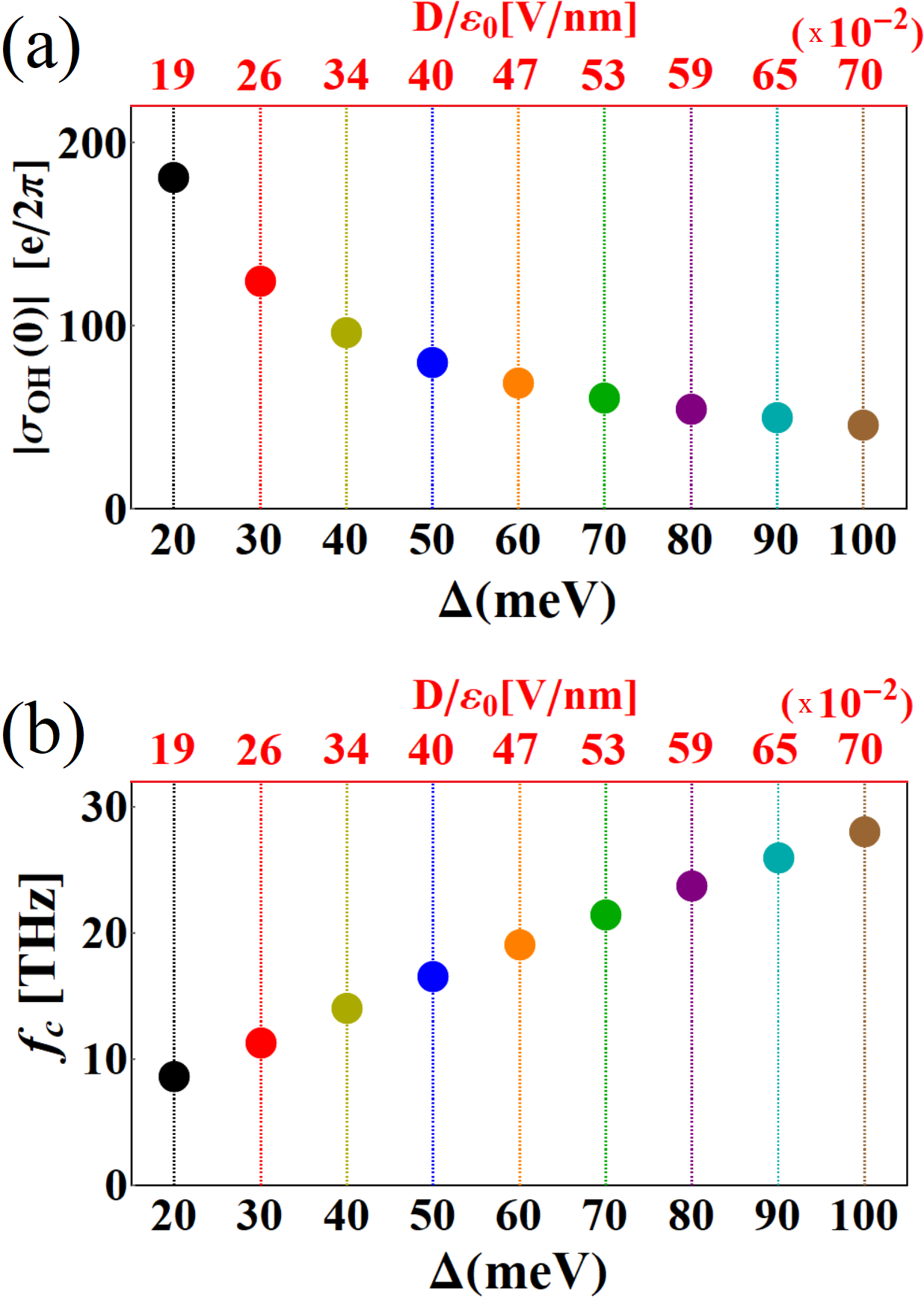}   
	\caption{Height of DC orbital Hall conductivity plateau $\left| \sigma_{\rm OH}(0)\right|$ and critical frequency $f_c$ in $\text{THz}$ as a function of $\Delta$. Here, we used $\eta=4$ meV. The red scale on the upper axis corresponds to the intensity of the displacement field associated with each value of layer asymmetry energy $\Delta$ obtained using the electrostatic model from Refs. \cite{McCann_2013-Eletrostatic, Nowack-PhysRevB.103.224426} for $E_{\rm F}=0$.}
	\label{OmegaC-Plateau_X_Delta}
\end{figure}

In Fig. \ref{Condutividade_X_HBarOmega} (a), we depict the behavior of the height of plateaus in the real part of AC OHC with $\hbar \omega/\Delta$ for various values of parameter $\Delta$ [see panel (c) for correspondence of the colors with $\Delta$]. The switch in the sign of plateaus in the real parts of OHC occurs at critical values $\hbar\omega_c/\Delta$, which depend on $\Delta$ [Fig. \ref{Condutividade_X_HBarOmega} (c)] and are represented by circles in the Fig. \ref{Condutividade_X_HBarOmega} (a). To complement the results of Fig. \ref{Condutividade_X_HBarOmega}, we show in Fig. \ref{OmegaC-Plateau_X_Delta} how the height of the real part of DC OHC plateau and the critical frequency $f_c$, in units of THz, vary with the value of $\Delta$. Notably, the behavior of critical frequency $f_c$ varies linearly with $\Delta$, i.e., it increases linearly with the layer asymmetry energy generated by electric bias. To connect the bilayer asymmetry energy $\Delta$ with the displacement field $D$ controlled in experiments, we employ the electrostatic model developed in Refs. \cite{McCann_2013-Eletrostatic, Nowack-PhysRevB.103.224426}. In this model one has $\Delta\approx e d_0 D/\varepsilon_0+\Lambda t_{\perp} \left(n_u-n_b\right)/n_0$, where the second term accounts for electrostatic screening effects. In these expressions, $d_0= 3.35 \angstrom$ is the inter-layer distance of BLG, $n_u$ and $n_b$ are the excess of carriers induced in the upper and bottom layers. $\Lambda\approx 1$ is the dimensionless screening parameter and $n_0$ is a constant with dimension of density. At $E_{\rm F}=0$, one obtains $n_{b(u)}/n_0 \approx \pm(\Delta/(4t_{\perp}))\ln \left(4t_{\perp}/|\Delta|\right)$ \cite{McCann_2013-Eletrostatic}. With these expressions, we can calculate the displacement field required in an experimental setup to generate the asymmetry energy $\Delta$, with the field intensities depicted along the upper red axis in Fig. \ref{OmegaC-Plateau_X_Delta}. For $\Delta \in [20-100]$ meV one demands $D/\varepsilon_0\in [0.19-0.70] \text{V/nm}$, that are typical values obtained in experiments \cite{Exp-DualgateBLG-PhysRevB.106.165134}. The critical frequency follows a linear behavior with the displacement field $f_c=\gamma D$. Using the data from the Fig. \ref{OmegaC-Plateau_X_Delta} (b), one obtains a fit for the numerical coefficient $\gamma=(40.5 \pm 0.3) \text{ THz nm/V}$. This demonstrates how one can adjust the behavior of OHC in the frequency domain in BLG using a tuning parameter that is easy to implement in experiments. 

Analogously, in Fig. \ref{Condutividade_X_HBarOmega} (b), we depict the behavior of the height of plateaus in the imaginary part of AC OHC with $\hbar \omega/\Delta$ for various values of parameter $\Delta$. As mentioned earlier, in contrast to the real part, the imaginary part of the OHC plateau maintains the negative sign throughout the entire frequency domain.  Furthermore, as we mentioned before, the real and imaginary parts of OHC are connected through causality constraints. One may notice that the general trend imposed by these constraints is followed by the results reported in Fig. \ref{Condutividade_X_HBarOmega}. See, for comparison, optical Hall conductivity in Ref. \cite{Wilton-PhysRevLett.119.147401}.

\section{Effect of weak magnetic field}

The time-reversal symmetry imposes constraints on the transport properties of the BLG. To complement our study in a situation where time-reversal symmetry is broken, we consider the introduction of a weak magnetic field perpendicular to the BLG ${\bf B}=B\hat{z}$. In the regime of a weak magnetic field, where Landau levels have not yet formed (for graphene $B\lessapprox 1.5T$ \cite{Us-Experiment-salvadorsanchez2022generation}), one can incorporate its effects through perturbation theory \cite{OMM-Kohn-PhysRev.115.1460, magField-Falde-NJPhys-2021, QNiu-PhysRevB.88.115140}. The correction to the electron's energy in the first order of the magnetic field is given by
\begin{eqnarray}
E^{B}_{n, {\bf q}_{\tau}}= E_{n, {\bf q}_{\tau}} - B \cdot \bra{u_{n, {\bf q}_{\tau}}} \hat{\mathbb{m}}^{z,{\rm tb}}_{{\bf q}_{\tau},{\rm BL}} \ket{u_{n, {\bf q}_{\tau}}}. \label{EB}
\end{eqnarray} 
The construction of the matrix $\hat{\mathbb{m}}^{z,{\rm tb}}_{{\bf q}_{\tau},{\rm BL}}$ is explained in detail in the Appendix \ref{AppendixA}. The correction in the periodic part of Bloch states is
\begin{eqnarray}
\ket{u^{B}_{n, {\bf q}_{\tau}}} = && \mathcal{N}\Bigg[ \ket{u_{n, {\bf q}_{\tau}}} \nonumber \\
&&-B\cdot \sum_{m \neq n} \frac{\bra{u_{m, {\bf q}_{\tau}}} \hat{\mathbb{m}}^{z,{\rm tb}}_{{\bf q}_{\tau},{\rm BL}} \ket{u_{n, {\bf q}_{\tau}}}}{\big( E_{n, {\bf q}_{\tau}} - E_{m, {\bf q}_{\tau}}\big)} \ket{u_{m, {\bf q}_{\tau}}} \Bigg], 
\label{uB}
\end{eqnarray}
where $\mathcal{N}$ is a normalization constant, i.e., $\mathcal{N}[\ket{v}]=(\braket{v|v})^{-1/2}\ket{v}$. Eqs. (\ref{EB}) and (\ref{uB}) encode the first-order perturbation theory in the external magnetic field. Higher-order perturbation theory was conducted in Ref. \cite{magField-Falde-NJPhys-2021}, revealing that for magnetic field intensities stronger than one used in the present work, corrections to the electronic density of states \cite{ModTheo-PhysRevLett.95.137204} may be important. However, for the weak magnetic field intensities employed here, these corrections do not play a significant role. To incorporate the influence of magnetic field on orbital Berry curvatures and conductivity, we perform the substitution $E_{n, {\bf q}_{\tau}}\rightarrow E^{B}_{n, {\bf q}_{\tau}}$ and $\ket{u_{n, {\bf q}_{\tau}}} \rightarrow \ket{u^{B}_{n, {\bf q}_{\tau}}}$ on Eqs. (\ref{sigmaOptcOH}, \ref{OBC}).

\begin{figure}
	\centering 
    \includegraphics[width=1\linewidth,clip]{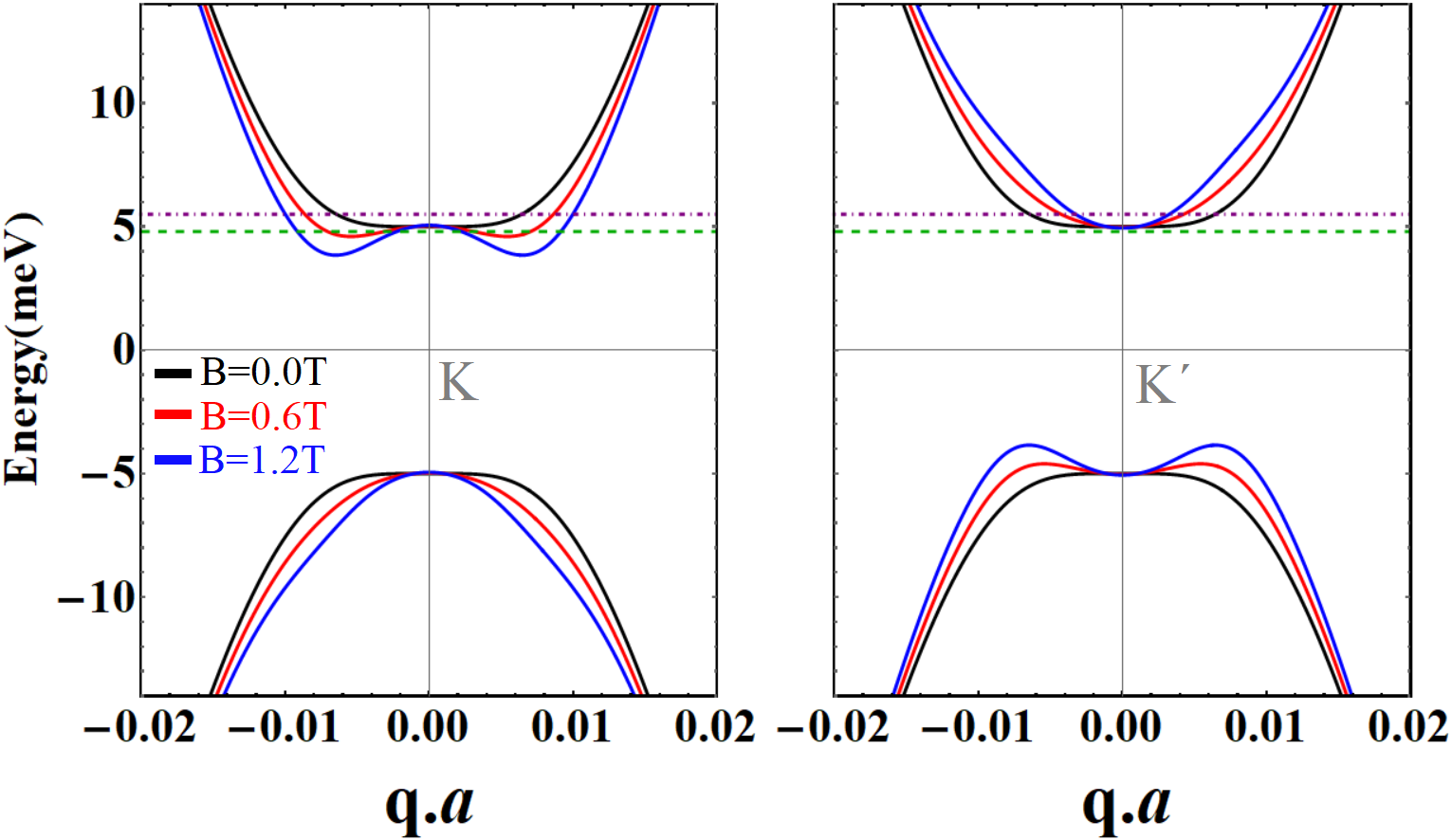}   
	\caption{The impact of a weak magnetic field on the top of the valence band $(1,v)$ and on the bottom of the conduction band $(1,c)$ in BLG with a bandgap $\Delta=10 \text{meV}$. The dashed green line (at $4.8 \text{meV}$) and the dot-dashed purple line (at $5.5 \text{meV}$) are the Fermi energy values discussed in the text.}
	\label{Espectra-MagField}
\end{figure} 

The breaking of time-reversal symmetry allows for the occurrence of a finite AC charge Hall effect governed by $\mathcal{J}_y(\omega)=\sigma_{{\rm H}}(\omega)\mathcal{E}_x(\omega)$. The AC Hall conductivity in the presence of a finite magnetic field is given by $\sigma_{\rm H}(\omega)=\sigma^{R}_{\rm H}(\omega)+i\sigma^{I}_{\rm H}(\omega)$ where \cite{AC-SHC-PhysRevLett.94.226601},
\begin{widetext}
\begin{eqnarray}
\sigma^{R/I}_{\rm H}(\omega)=\frac{e^2}{\hbar}\sum_n \sum_{\tau=\pm 1}\int \frac{d^2q}{(2\pi)^2}\Omega^{R/I}_{{\rm H},n} ({\bf q}_{\tau},\omega) \Theta(E_{\rm F}-E^B_{n,{\bf q}_{\tau}}), \label{sigmaOptcH}
\end{eqnarray}
and,
\begin{eqnarray}
\frac{\Omega^{R/I}_{{\rm H},n}({\bf q}_{\tau},\omega)}{2\hbar^2}&=&\frac{1}{2}\sum_{m\neq n} \begin{Bmatrix}
  \text{Im} \\ -\text{Re} \\\end{Bmatrix} \Bigg[\frac{1}{(E^B_{n,{\bf q}_{\tau}}-E^B_{m,{\bf q}_{\tau}})}\Bigg(\frac{\bra{u^B_{n,{\bf q}_{\tau}}}\hat{v}_x({\bf q}_{\tau})\ket{u^B_{m,{\bf q}_{\tau}}} \bra{u^B_{m,{\bf q}_{\tau}}}\hat{v}_y({\bf q}_{\tau})\ket{u^B_{n,{\bf q}_{\tau}}}}{(E^B_{n,{\bf q}_{\tau}}-E^B_{m,{\bf q}_{\tau}}+\hbar \omega+i\eta)} + (n\leftrightarrow m)\Bigg)\Bigg]. \label{OmegaH}
\end{eqnarray}
\end{widetext}
Similar to what is done for AC OHC in the presence of a perpendicular magnetic field, one employs first-order perturbation theory to correct energy and vectors in the calculation of AC Hall conductivity in Eqs. (\ref{sigmaOptcH}, \ref{OmegaH}).

The influence of the weak magnetic field in Eqs. (\ref{EB}, \ref{uB}) is directly proportional to the matrix $\hat{\mathbb{m}}^{z,{\rm tb}}_{{\bf q}_{\tau},{\rm BL}}$ and becomes more pronounced for narrower bandgaps $\Delta$. In this section, we explore the case of $\Delta=10 \ \text{meV}$. In Fig. \ref{Espectra-MagField}, we depict the electronic spectra near the band edges of BLG [states $(1,c)$ and $(1,v)$ of Fig. \ref{BLGSpectra_mz} (a)] for various values of $B$. Due to the break in time-reversal symmetry, the energy branches at valley ${\bf K}$ are no longer degenerate with those at valley ${\bf K'}$ for $B\neq 0$. The distortion in the energy spectra alters the Fermi-Dirac distributions $\Theta (E_{\rm F}-E^{B}_{n,{\bf q}_{\tau}})$ in Eqs. (\ref{sigmaOptcOH}, \ref{sigmaOptcH}), representing the primary quantitative effect of the magnetic field. This results in significant changes in the transport properties when the Fermi energy is positioned near the band edges, as indicated by the purple and green lines in Fig. \ref{Espectra-MagField}. 

\begin{figure}
	\centering
	 \includegraphics[width=1.0\linewidth,clip]{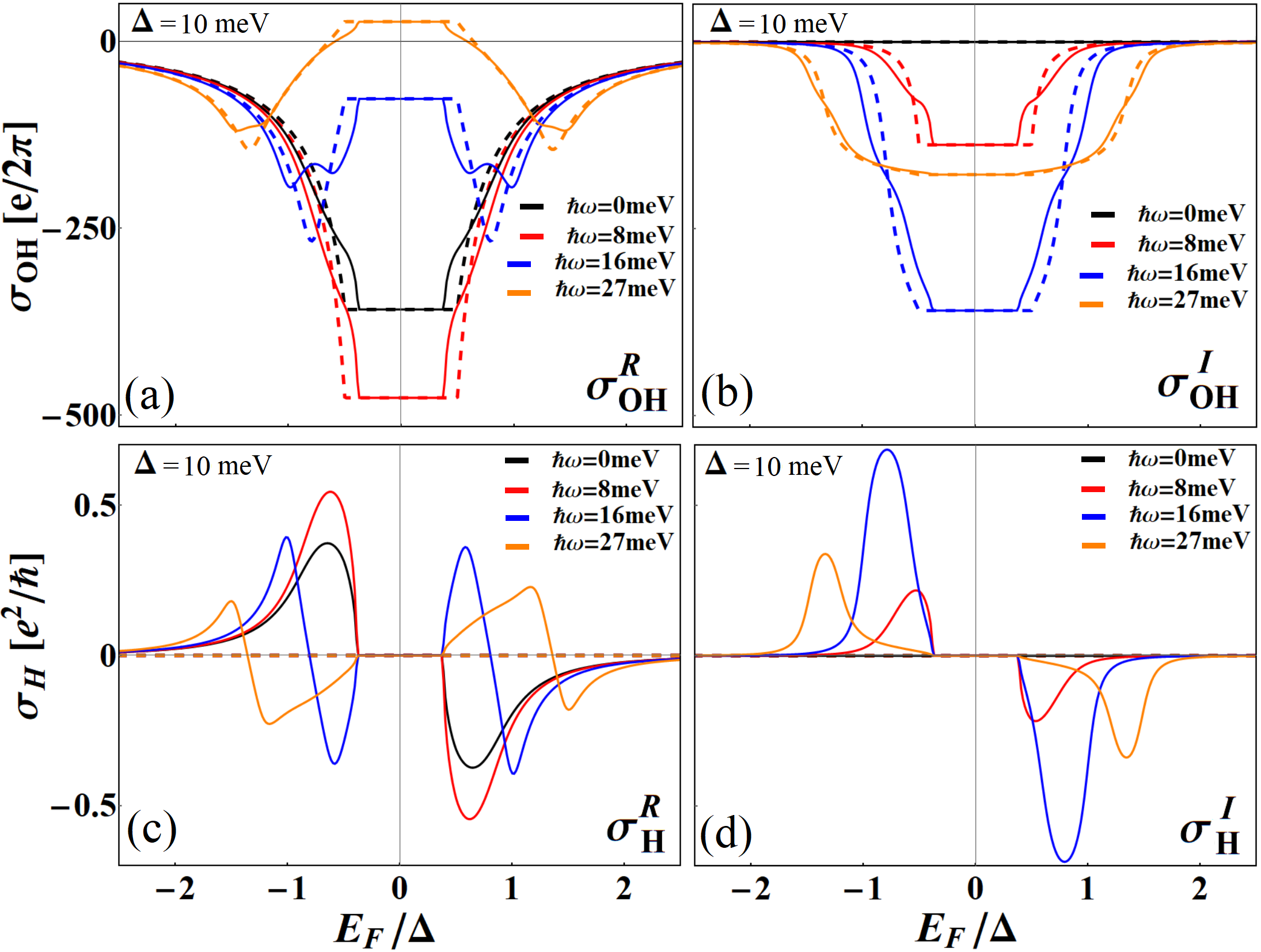}   
	\caption{(a) Real and (b) Imaginary parts of OHC as a function of $E_{\rm F}/\Delta$ for different values of frequency. (c) Real and (d) Imaginary parts of charge Hall conductivity as a function of $E_{\rm F}/\Delta$ for different values of frequency. We used $\Delta=10 \text{meV}$ and set $\eta=2$ meV. Solid (Dashed) lines show the results for an applied magnetic field $B=1.2$T ($B=0$T).}
	\label{OHC_X_EF-BFinite}
\end{figure} 

In Fig. \ref{OHC_X_EF-BFinite} (a), we show the real part of the AC OHC as a function of Fermi energy for $\Delta=10 \text{ meV}$ and different values of frequency. The solid curve represents the results in the presence of an external magnetic field $B=1.2 \text{ T}$, and the dashed curves represent the results in the absence of a magnetic field $B=0.0 \text{ T}$. In panel (b) of this figure, we show analogous results, but for the imaginary part of the AC OHC. The OHC is affected by the magnetic field only in the metallic regime when the Fermi energy is located close to the band edges. When the Fermi energy lies inside the insulating bandgap, the height of the OHC plateau is not modified by the presence of the magnetic field. In Fig. \ref{OHC_X_EF-BFinite} (c) and (d), we present the results of the charge Hall conductivity. One notices that the gapped BLG under the influence of an external magnetic field has a finite AC charge Hall conductivity when the Fermi energy crosses electronic bands.

\begin{figure}[h]
	\centering
	 \includegraphics[width=0.85\linewidth,clip]{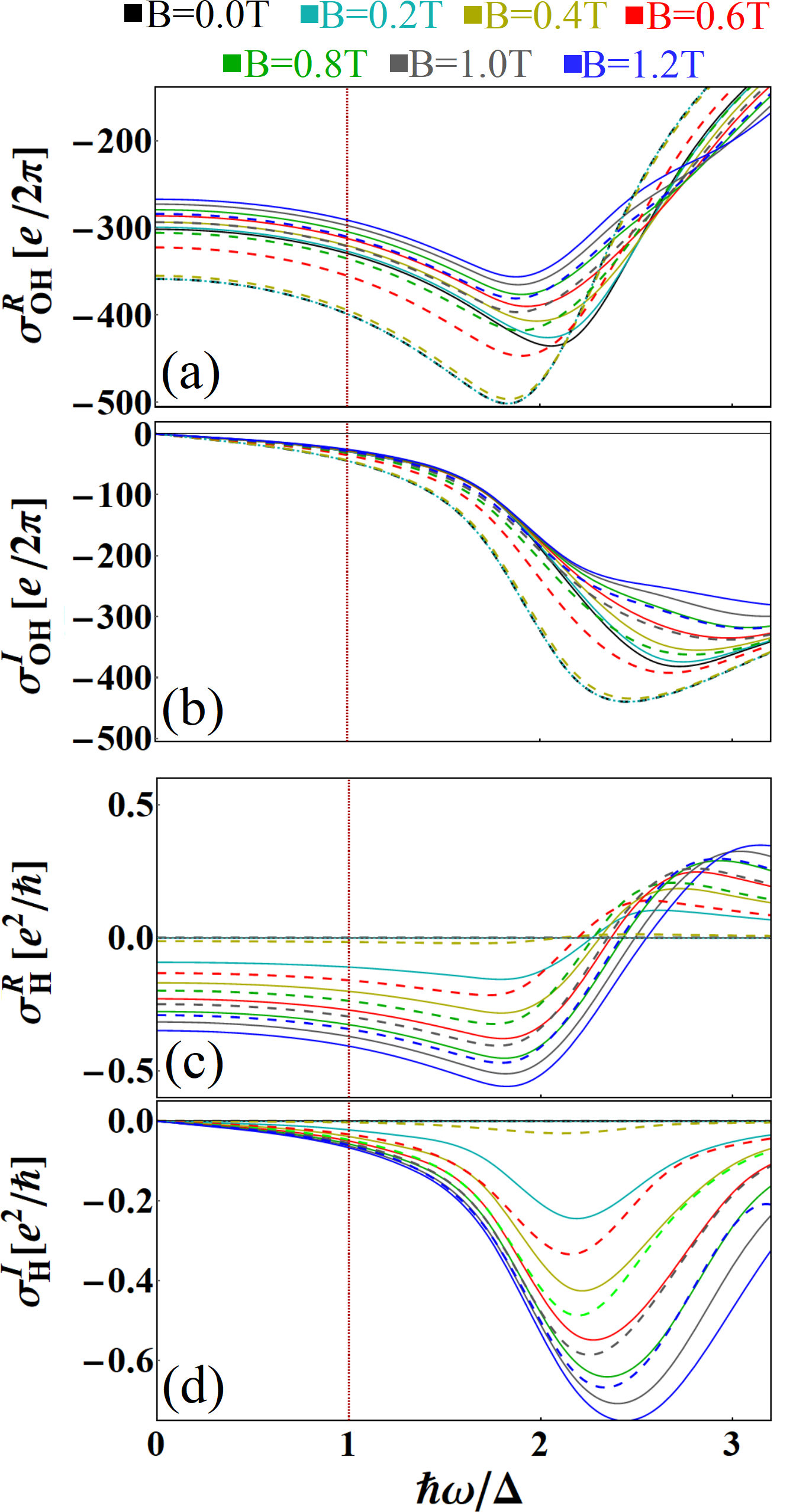}   
	\caption{(a) Real and (b) imaginary parts of the AC OHC as functions of $\hbar\omega/\Delta$ for various values of the magnetic field. Panels (c) and (d) show analogous results but for AC Hall conductivity. Solid curves represent the results for $E_{\rm F}=5.5 \text{meV}$ and the dashed curves represent the results for $E_{\rm F}=4.8 \text{meV}$. Here, we set $\Delta=10 \ \text{meV}$ and $\eta=2$ meV.}
	\label{Cond-Omega-B}
\end{figure}  

In Fig. \ref{Cond-Omega-B}, we show the behavior of orbital and charge AC Hall conductivities in the presence of weak magnetic fields for two different values of Fermi energy close to the bottom of the conduction band [dashed green and dot-dashed purple lines in Fig. \ref{Espectra-MagField}]. The magnetic field modifies the OHC through the entire frequency range when the Fermi energy crosses the energy bands, i.e., out of the charge neutrality (metallic) situation [Fig. \ref{Cond-Omega-B} (a) and (b)]. As discussed above, the magnetic field does not influence the height of the OHC plateaus when $E_{\rm F}$ is situated within the band gap. Furthermore, the magnetic field allows $\sigma^{R/I}_{H}(\omega)\neq 0$ in Eq. (\ref{sigmaOptcH}) and then a finite Hall conductivity appears as illustrated in Fig. \ref{Cond-Omega-B} (c) and (d). In the metallic BLG in the presence of a magnetic field, both charge and orbital angular momentum accumulate at the edges of the sample.

\section{Final remarks and conclusion}

Our results suggest that an AC orbital Hall current can be generated in BLG, and its intensity may be controlled by applying a perpendicular electric field. Recent experiments in bilayers of metallic thin films have observed the dynamic generation of orbital currents in ultra-fast regimes and their conversion into charge currents, resulting in the emission of terahertz radiation \cite{Fert-xu2023orbitronics, Go-Thz-emission-10.1038/s41565-023-01470-8, IOHE-THzEmission-Kumar2023-NatComm, IOHE-THz-s41535-023-00559-6}. Furthermore, Refs. \cite{IOHE-THzEmission-Kumar2023-NatComm, IOHE-THz-s41535-023-00559-6} observed the terahertz emission signal and attributed it to the inverse OHE, in analogy to the effect driven by the spin counterpart \cite{Oppeneer-10.1038/nnano.2013.43}. The recent progress in spintronics within the terahertz regime has sparked numerous experimental and theoretical studies on time-dependent spin dynamics \cite{Bechara-THz-2017, Bechara-PhysRevB.92.220410, Brouwer-PhysRevB.106.144423, Brouwer-PhysRevB.104.024415, Silva-PhysRevLett.113.157204, Walowski-10.1063_1.4958846}. The results presented here could prove beneficial for the development of analogous orbitronic counterparts utilizing 2D materials.  

It is worth mentioning that the electrically controllable band gap of BLG is a pivotal factor in our study. Controlling the parameter $\Delta$ allows one to alter the intensity and shape of the orbital Hall conductivity curve across the frequency domain [see Figs. \ref{Condutividade_X_HBarOmega} and \ref{OmegaC-Plateau_X_Delta}]. In the case of monolayer graphene, this would not be possible because it does not respond to the perpendicular electric field. Transition metal dichalcogenides exhibit a rigid band gap and are only weakly influenced by this type of perturbation.

We conclude this paper by noting that the results reported here would be more readily observable in very narrow gaps (very weak perpendicular electric/displacement field). In such a situation, the orbital Hall conductivity is enhanced due to the high curvature in the Bloch bands. Additionally, the relevant frequency range for orbital transport is shifted to slower regimes, which should be easier to probe. This scenario can be replicated in experiments, as the current technology allows for the production of BLG with narrow gaps and its interaction with light fields \cite{Koppens-Sience-2023}.   

\begin{acknowledgments}
	We acknowledge CNPq/Brazil, CAPES/Brazil, FAPERJ/Brazil,  INCT Nanocarbono for financial support. TGR acknowledges funding from FCT-Portugal through Grant No. 2022.07471.CEECIND/CP1718/CT0001 (\url{https://doi.org/10.54499/2022.07471.CEECIND/CP1718/CT0001}). W.K.-K. acknowledges the Laboratory Directed Research and Development program of Los Alamos National Laboratory under Projects No. 20220273ER and 20240037DR. T.P.C. acknowledges Roberto B. Muniz for fruitful discussions. 
\end{acknowledgments} 


\appendix
\section{The orbital magnetic moment of Bloch-states \label{AppendixA}}
\subsection{The formalism for an isolated set of Bloch bands}
The Bloch states of electrons in solids exhibit an inherent orbital magnetic moment \cite{OMM-QNiu-PhysRevB.53.7010, OMM-Kohn-PhysRev.115.1460}. In the case of coupled Bloch bands, this orbital magnetic moment assumes a non-Abelian (matricial) structure \cite{Culcer-PhysRevB.72.085110}. This non-Abelian nature of the orbital magnetic moment is essential for describing orbital transport in centrosymmetric systems, like unbiased bilayers of 2H transition metal dichalcogenides \cite{Cysne-Bhowal-Vignale-Rappoport-PhysRevB.105.195421}. To construct the orbital magnetic moment matrix, we follow Ref. \cite{Culcer-PhysRevB.72.085110} and consider a subset of Hilbert space of dimension $N_{\rm sub}<\text{dim}[\hat{H}({\bf k})]$ isolated from the rest of the Hamiltonian's energy spectrum by an energy gap. In the case of the BLG that we are studying here, there are two subsets consisting of conduction and valence bands separated by an energy gap proportional to $\Delta$. 
The Bloch states inside the subset have periodic parts represented by $\ket{u_{\bar{n},{\bf k}}}$ and energies $E_{\bar{n},{\bf k}}$, where $\bar{n}=1,2,...,N_{\rm sub}$. The elements of the orbital magnetic moment matrix are given by
\begin{eqnarray}
&&m_{\bar{n},\bar{m}}^{z,u}({\bf k})=-i\frac{e}{2\hbar}\nonumber \\
&& \bra{\vec{\nabla}_{{\bf k}} u_{\bar{n},{\bf k}}} \boldsymbol{\times} \left[ \hat{H}({\bf k})-\left(\frac{E_{\bar{n},{\bf k}}+E_{\bar{m},{\bf k}}}{2}  \right)\hat{\mathbb{1}} \right] \ket{\vec{\nabla}_{{\bf k}} u_{\bar{m},{\bf k}}}. \nonumber \\
\label{MMDegenerate}
\end{eqnarray}
where, $\vec{\nabla}_{\bf k}=\hat{x}\partial/\partial k_x+\hat{y}\partial/\partial k_y$ and $\boldsymbol{\times}$ represents the cross-product. The $N_{\rm sub}$-dimensional orbital magnetic moment matrix can be cast as \cite{Culcer-PhysRevB.72.085110}, 
\begin{eqnarray}
\hat{\mathbb{m}}^{z,u}_{{\bf k}}= \begin{bmatrix}
m_{1,1}^{z,u}({\bf k}) & m_{1,2}^{z,u}({\bf k}) & \ldots & m_{1,N_{\rm sub}}^{z,u}({\bf k})\\
m_{2,1}^{z,u}({\bf k})  & m_{2,2}^{z,u}({\bf k}) & \ldots & m_{2,N_{\rm sub}}^{z,u}({\bf k}) \\
\vdots & \vdots & \ddots  & \vdots \\
m_{N_{\rm sub},1}^{z,u}({\bf k}) & m_{N_{\rm sub},2}^{z,u}({\bf k}) & \ldots & m_{N_{\rm sub},N_{\rm sub}}^{z,u}({\bf k}) 
\end{bmatrix}.\nonumber \\ \label{MMDegenerateMatrix}
\end{eqnarray}
The matrix in Eq. (\ref{MMDegenerateMatrix}) is written in the basis of Hamiltonian's eigenstates $\beta_u=\{u_1; u_2; ... ; u_{N_{\rm sub}} \}$. After obtaining the above matrix, we proceed by defining a unitary transformation $\hat{U}({\bf k}): \beta_u \rightarrow \beta_{\rm tb}$ and then applying it to Eq. (\ref{MMDegenerateMatrix}) in order to obtain the orbital magnetic moment operator within a tight-binding basis \cite{Cysne-Bhowal-Vignale-Rappoport-PhysRevB.105.195421}
\begin{eqnarray}
\hat{\mathbb{m}}^{z,{\rm tb}}_{{\bf k}}= \hat{U}({\bf k}) \hat{\mathbb{m}}^{z,u}_{{\bf k}} \hat{U}^{\dagger}({\bf k}). \label{mtb} 
\end{eqnarray}
Following Refs. \cite{Bhowal-Vignale-PhysRevB.103.195309, Cysne-Bhowal-Vignale-Rappoport-PhysRevB.105.195421}, one uses the matrix $\hat{\mathbb{m}}^{z,{\rm tb}}_{{\bf k}}$ to define the OAM operator of the electronic Bloch states
\begin{eqnarray} 
\hat{L}_z({\bf k})=-(\hbar/\mu_Bg_L)\hat{\mathbb{m}}^{z,{\rm tb}}_{{\bf k}}. \label{LzBloch}
\end{eqnarray}
In multiorbital systems, this operator includes contributions from both intersite and intrasite motions of electrons \cite{ISouza-modTheo-PhysRevB.77.054438, FXuan-modTheo-PhysRevResearch.2.033256}. It is employed in the Kubo formula [Eq. (\ref{sigmaOptcOH})] for computing the OHC. 

\subsection{Application to gapped bilayer graphene}
In the case of the BLG [Eq. (\ref{HBGB})], the conduction (c) and valence (v) bands each form a subspace of dimension two, and they are separated by a bandgap proportional to $\Delta$, as illustrated in Fig. \ref{BLGSpectra_mz} (a). The construction of the orbital magnetic moment matrix from Eq. (\ref{MMDegenerateMatrix}) is then applied separately to the conduction and valence 2-dimensional subspaces \cite{Cysne-Bhowal-Vignale-Rappoport-PhysRevB.105.195421, Culcer-PhysRevB.72.085110}. In the basis of Bloch eigenstates $\beta^{\rm BL}_u=\{(2,v), \ (1,v), \ (1,c), \ (2,c) \}$ it can be cast as
\begin{eqnarray}
\hat{\mathbb{m}}^{z,u}_{{\bf q}_{\tau},{\rm BL}}= \begin{bmatrix}
m^{(2,v)}_{{\bf q}_{\tau}} & m^{\text{v,od}}_{{\bf q}_{\tau}} & 0 & 0\\
\big(m^{\text{v,od}}_{{\bf q}_{\tau}}\big)^*  & m^{(1,v)}_{{\bf q}_{\tau}} & 0 & 0 \\
0 & 0 & m^{(1,c)}_{{\bf q}_{\tau}}  & m^{\text{c,od}}_{{\bf q}_{\tau}} \\
0 & 0 & \big(m^{\text{c,od}}_{{\bf q}_{\tau}}\big)^*  & m^{(2,c)}_{{\bf q}_{\tau}} 
\end{bmatrix}. \nonumber \\ \label{MMDegenerateBLgraphene}
\end{eqnarray}
The finite $\Delta$ breaks the spatial inversion symmetry of the bilayer system, leading to non-zero diagonal elements in the matrix [Fig. \ref{BLGSpectra_mz} (b)]. The off-diagonal matrix elements are not constrained by spatial inversion symmetry and can appear even in centrosymmetric models, such as in an unbiased bilayer of 2H transition metal dichalcogenides \cite{Cysne-Bhowal-Vignale-Rappoport-PhysRevB.105.195421}. We then define the unitary transformation $\hat{U}_{\rm BL}({\bf q}_{\tau}): \beta^{\rm BL}_u \rightarrow \beta^{\rm BL}_{tb} (\tau {\bf K})$ from Bloch eigenstates to the tight-binding basis of BLG defined in Sec. \ref{Sec:model}. The Bloch state orbital magnetic moment matrix in tight-binding basis is given by $\hat{\mathbb{m}}^{z,{\rm tb}}_{{\bf q}_{\tau},{\rm BL}}= \hat{U}({\bf q}_{\tau}) \hat{\mathbb{m}}^{z,u}_{{\bf q}_{\tau},{\rm BL}}\hat{U}^{\dagger}({\bf q}_{\tau})$. We then use Eq. (\ref{LzBloch}) and apply it to the Kubo formula [Eq. (\ref{sigmaOptcOH})] to calculate the orbital Hall conductivity.

It is worth mentioning that the electronic structure of graphene is dominated by $p_z$ orbitals, which do not possess intra-atomic OAM. In the context of the BLG Hamiltonian [Eq. (\ref{HBGB})] examined in this work, the contribution of OAM in the Eq. (\ref{MMDegenerateBLgraphene}) arises exclusively from the intersite movement of electrons \cite{Bhowal-Vignale-PhysRevB.103.195309, ISouza-modTheo-PhysRevB.77.054438, FXuan-modTheo-PhysRevResearch.2.033256}.


\begin{thebibliography}{93}%
\makeatletter
\providecommand \@ifxundefined [1]{%
 \@ifx{#1\undefined}
}%
\providecommand \@ifnum [1]{%
 \ifnum #1\expandafter \@firstoftwo
 \else \expandafter \@secondoftwo
 \fi
}%
\providecommand \@ifx [1]{%
 \ifx #1\expandafter \@firstoftwo
 \else \expandafter \@secondoftwo
 \fi
}%
\providecommand \natexlab [1]{#1}%
\providecommand \enquote  [1]{``#1''}%
\providecommand \bibnamefont  [1]{#1}%
\providecommand \bibfnamefont [1]{#1}%
\providecommand \citenamefont [1]{#1}%
\providecommand \href@noop [0]{\@secondoftwo}%
\providecommand \href [0]{\begingroup \@sanitize@url \@href}%
\providecommand \@href[1]{\@@startlink{#1}\@@href}%
\providecommand \@@href[1]{\endgroup#1\@@endlink}%
\providecommand \@sanitize@url [0]{\catcode `\\12\catcode `\$12\catcode
  `\&12\catcode `\#12\catcode `\^12\catcode `\_12\catcode `\%12\relax}%
\providecommand \@@startlink[1]{}%
\providecommand \@@endlink[0]{}%
\providecommand \url  [0]{\begingroup\@sanitize@url \@url }%
\providecommand \@url [1]{\endgroup\@href {#1}{\urlprefix }}%
\providecommand \urlprefix  [0]{URL }%
\providecommand \Eprint [0]{\href }%
\providecommand \doibase [0]{https://doi.org/}%
\providecommand \selectlanguage [0]{\@gobble}%
\providecommand \bibinfo  [0]{\@secondoftwo}%
\providecommand \bibfield  [0]{\@secondoftwo}%
\providecommand \translation [1]{[#1]}%
\providecommand \BibitemOpen [0]{}%
\providecommand \bibitemStop [0]{}%
\providecommand \bibitemNoStop [0]{.\EOS\space}%
\providecommand \EOS [0]{\spacefactor3000\relax}%
\providecommand \BibitemShut  [1]{\csname bibitem#1\endcsname}%
\let\auto@bib@innerbib\@empty
\bibitem [{\citenamefont {Bernevig}\ \emph {et~al.}(2005)\citenamefont
  {Bernevig}, \citenamefont {Hughes},\ and\ \citenamefont
  {Zhang}}]{Bernevig-PhysRevLett.95.066601}%
  \BibitemOpen
  \bibfield  {author} {\bibinfo {author} {\bibfnamefont {B.~A.}\ \bibnamefont
  {Bernevig}}, \bibinfo {author} {\bibfnamefont {T.~L.}\ \bibnamefont
  {Hughes}},\ and\ \bibinfo {author} {\bibfnamefont {S.-C.}\ \bibnamefont
  {Zhang}},\ }\bibfield  {title} {\bibinfo {title} {Orbitronics: The intrinsic
  orbital current in $p$-doped silicon},\ }\href
  {https://doi.org/10.1103/PhysRevLett.95.066601} {\bibfield  {journal}
  {\bibinfo  {journal} {Phys. Rev. Lett.}\ }\textbf {\bibinfo {volume} {95}},\
  \bibinfo {pages} {066601} (\bibinfo {year} {2005})}\BibitemShut {NoStop}%
\bibitem [{\citenamefont {Choi}\ \emph {et~al.}(2023)\citenamefont {Choi},
  \citenamefont {Jo}, \citenamefont {Ko}, \citenamefont {Go}, \citenamefont
  {Kim}, \citenamefont {Park}, \citenamefont {Kim}, \citenamefont {Min},
  \citenamefont {Choi},\ and\ \citenamefont
  {Lee}}]{Nature-OrbitalHall-Hyun-Woo}%
  \BibitemOpen
  \bibfield  {author} {\bibinfo {author} {\bibfnamefont {Y.-G.}\ \bibnamefont
  {Choi}}, \bibinfo {author} {\bibfnamefont {D.}~\bibnamefont {Jo}}, \bibinfo
  {author} {\bibfnamefont {K.-H.}\ \bibnamefont {Ko}}, \bibinfo {author}
  {\bibfnamefont {D.}~\bibnamefont {Go}}, \bibinfo {author} {\bibfnamefont
  {K.-H.}\ \bibnamefont {Kim}}, \bibinfo {author} {\bibfnamefont {H.~G.}\
  \bibnamefont {Park}}, \bibinfo {author} {\bibfnamefont {C.}~\bibnamefont
  {Kim}}, \bibinfo {author} {\bibfnamefont {B.-C.}\ \bibnamefont {Min}},
  \bibinfo {author} {\bibfnamefont {G.-M.}\ \bibnamefont {Choi}},\ and\
  \bibinfo {author} {\bibfnamefont {H.-W.}\ \bibnamefont {Lee}},\ }\bibfield
  {title} {\bibinfo {title} {Observation of the orbital hall effect in a light
  metal ti},\ }\href {https://doi.org/10.1038/s41586-023-06101-9} {\bibfield
  {journal} {\bibinfo  {journal} {Nature}\ }\textbf {\bibinfo {volume} {619}},\
  \bibinfo {pages} {52} (\bibinfo {year} {2023})}\BibitemShut {NoStop}%
\bibitem [{\citenamefont {Lyalin}\ \emph {et~al.}(2023)\citenamefont {Lyalin},
  \citenamefont {Alikhah}, \citenamefont {Berritta}, \citenamefont {Oppeneer},\
  and\ \citenamefont {Kawakami}}]{DetectionOHE-PhysRevLett.131.156702}%
  \BibitemOpen
  \bibfield  {author} {\bibinfo {author} {\bibfnamefont {I.}~\bibnamefont
  {Lyalin}}, \bibinfo {author} {\bibfnamefont {S.}~\bibnamefont {Alikhah}},
  \bibinfo {author} {\bibfnamefont {M.}~\bibnamefont {Berritta}}, \bibinfo
  {author} {\bibfnamefont {P.~M.}\ \bibnamefont {Oppeneer}},\ and\ \bibinfo
  {author} {\bibfnamefont {R.~K.}\ \bibnamefont {Kawakami}},\ }\bibfield
  {title} {\bibinfo {title} {Magneto-optical detection of the orbital hall
  effect in chromium},\ }\href {https://doi.org/10.1103/PhysRevLett.131.156702}
  {\bibfield  {journal} {\bibinfo  {journal} {Phys. Rev. Lett.}\ }\textbf
  {\bibinfo {volume} {131}},\ \bibinfo {pages} {156702} (\bibinfo {year}
  {2023})}\BibitemShut {NoStop}%
\bibitem [{\citenamefont {Sala}\ \emph {et~al.}(2023)\citenamefont {Sala},
  \citenamefont {Wang}, \citenamefont {Legrand},\ and\ \citenamefont
  {Gambardella}}]{DetectionOHE-PhysRevLett.131.156703}%
  \BibitemOpen
  \bibfield  {author} {\bibinfo {author} {\bibfnamefont {G.}~\bibnamefont
  {Sala}}, \bibinfo {author} {\bibfnamefont {H.}~\bibnamefont {Wang}}, \bibinfo
  {author} {\bibfnamefont {W.}~\bibnamefont {Legrand}},\ and\ \bibinfo {author}
  {\bibfnamefont {P.}~\bibnamefont {Gambardella}},\ }\bibfield  {title}
  {\bibinfo {title} {Orbital hanle magnetoresistance in a $3d$ transition
  metal},\ }\href {https://doi.org/10.1103/PhysRevLett.131.156703} {\bibfield
  {journal} {\bibinfo  {journal} {Phys. Rev. Lett.}\ }\textbf {\bibinfo
  {volume} {131}},\ \bibinfo {pages} {156703} (\bibinfo {year}
  {2023})}\BibitemShut {NoStop}%
\bibitem [{\citenamefont {Go}\ \emph {et~al.}(2021)\citenamefont {Go},
  \citenamefont {Jo}, \citenamefont {Lee}, \citenamefont {Kläui},\ and\
  \citenamefont {Mokrousov}}]{Go-EPL-MiniReview_2021}%
  \BibitemOpen
  \bibfield  {author} {\bibinfo {author} {\bibfnamefont {D.}~\bibnamefont
  {Go}}, \bibinfo {author} {\bibfnamefont {D.}~\bibnamefont {Jo}}, \bibinfo
  {author} {\bibfnamefont {H.-W.}\ \bibnamefont {Lee}}, \bibinfo {author}
  {\bibfnamefont {M.}~\bibnamefont {Kläui}},\ and\ \bibinfo {author}
  {\bibfnamefont {Y.}~\bibnamefont {Mokrousov}},\ }\bibfield  {title} {\bibinfo
  {title} {Orbitronics: Orbital currents in solids},\ }\href
  {https://doi.org/10.1209/0295-5075/ac2653} {\bibfield  {journal} {\bibinfo
  {journal} {Europhysics Letters}\ }\textbf {\bibinfo {volume} {135}},\
  \bibinfo {pages} {37001} (\bibinfo {year} {2021})}\BibitemShut {NoStop}%
\bibitem [{\citenamefont {Urazhdin}(2023)}]{urazhdin2023symmetry}%
  \BibitemOpen
  \bibfield  {author} {\bibinfo {author} {\bibfnamefont {S.}~\bibnamefont
  {Urazhdin}},\ }\bibfield  {title} {\bibinfo {title} {Symmetry constraints on
  orbital transport in solids},\ }\href
  {https://doi.org/10.1103/PhysRevB.108.L180404} {\bibfield  {journal}
  {\bibinfo  {journal} {Phys. Rev. B}\ }\textbf {\bibinfo {volume} {108}},\
  \bibinfo {pages} {L180404} (\bibinfo {year} {2023})}\BibitemShut {NoStop}%
\bibitem [{\citenamefont {Go}\ \emph {et~al.}(2018)\citenamefont {Go},
  \citenamefont {Jo}, \citenamefont {Kim},\ and\ \citenamefont
  {Lee}}]{Go-Textures-PhysRevLett.121.086602}%
  \BibitemOpen
  \bibfield  {author} {\bibinfo {author} {\bibfnamefont {D.}~\bibnamefont
  {Go}}, \bibinfo {author} {\bibfnamefont {D.}~\bibnamefont {Jo}}, \bibinfo
  {author} {\bibfnamefont {C.}~\bibnamefont {Kim}},\ and\ \bibinfo {author}
  {\bibfnamefont {H.-W.}\ \bibnamefont {Lee}},\ }\bibfield  {title} {\bibinfo
  {title} {Intrinsic spin and orbital hall effects from orbital texture},\
  }\href {https://doi.org/10.1103/PhysRevLett.121.086602} {\bibfield  {journal}
  {\bibinfo  {journal} {Phys. Rev. Lett.}\ }\textbf {\bibinfo {volume} {121}},\
  \bibinfo {pages} {086602} (\bibinfo {year} {2018})}\BibitemShut {NoStop}%
\bibitem [{\citenamefont {Sala}\ and\ \citenamefont
  {Gambardella}(2022)}]{Gambardella-PhysRevResearch.4.033037}%
  \BibitemOpen
  \bibfield  {author} {\bibinfo {author} {\bibfnamefont {G.}~\bibnamefont
  {Sala}}\ and\ \bibinfo {author} {\bibfnamefont {P.}~\bibnamefont
  {Gambardella}},\ }\bibfield  {title} {\bibinfo {title} {Giant orbital hall
  effect and orbital-to-spin conversion in $3d$, $5d$, and $4f$ metallic
  heterostructures},\ }\href {https://doi.org/10.1103/PhysRevResearch.4.033037}
  {\bibfield  {journal} {\bibinfo  {journal} {Phys. Rev. Res.}\ }\textbf
  {\bibinfo {volume} {4}},\ \bibinfo {pages} {033037} (\bibinfo {year}
  {2022})}\BibitemShut {NoStop}%
\bibitem [{\citenamefont {Atencia}\ and\ \citenamefont
  {Culcer}(2023)}]{atencia2023nonconservation}%
  \BibitemOpen
  \bibfield  {author} {\bibinfo {author} {\bibfnamefont {R.~B.}\ \bibnamefont
  {Atencia}}\ and\ \bibinfo {author} {\bibfnamefont {D.}~\bibnamefont
  {Culcer}},\ }\href@noop {} {\bibinfo {title} {Non-conservation of the orbital
  moment of bloch electrons in an electric field}} (\bibinfo {year} {2023}),\
  \Eprint {https://arxiv.org/abs/2311.12108} {arXiv:2311.12108
  [cond-mat.mes-hall]} \BibitemShut {NoStop}%
\bibitem [{\citenamefont {Bose}\ \emph {et~al.}(2023)\citenamefont {Bose},
  \citenamefont {Kammerbauer}, \citenamefont {Gupta}, \citenamefont {Go},
  \citenamefont {Mokrousov}, \citenamefont {Jakob},\ and\ \citenamefont
  {Kl\"aui}}]{ArnabBose-PhysRevB.107.134423}%
  \BibitemOpen
  \bibfield  {author} {\bibinfo {author} {\bibfnamefont {A.}~\bibnamefont
  {Bose}}, \bibinfo {author} {\bibfnamefont {F.}~\bibnamefont {Kammerbauer}},
  \bibinfo {author} {\bibfnamefont {R.}~\bibnamefont {Gupta}}, \bibinfo
  {author} {\bibfnamefont {D.}~\bibnamefont {Go}}, \bibinfo {author}
  {\bibfnamefont {Y.}~\bibnamefont {Mokrousov}}, \bibinfo {author}
  {\bibfnamefont {G.}~\bibnamefont {Jakob}},\ and\ \bibinfo {author}
  {\bibfnamefont {M.}~\bibnamefont {Kl\"aui}},\ }\bibfield  {title} {\bibinfo
  {title} {Detection of long-range orbital-hall torques},\ }\href
  {https://doi.org/10.1103/PhysRevB.107.134423} {\bibfield  {journal} {\bibinfo
   {journal} {Phys. Rev. B}\ }\textbf {\bibinfo {volume} {107}},\ \bibinfo
  {pages} {134423} (\bibinfo {year} {2023})}\BibitemShut {NoStop}%
\bibitem [{\citenamefont {Dowinton}\ and\ \citenamefont
  {Bahramy}(2022)}]{PhysRevB.105.235142}%
  \BibitemOpen
  \bibfield  {author} {\bibinfo {author} {\bibfnamefont {O.}~\bibnamefont
  {Dowinton}}\ and\ \bibinfo {author} {\bibfnamefont {M.~S.}\ \bibnamefont
  {Bahramy}},\ }\bibfield  {title} {\bibinfo {title} {Orbital angular momentum
  driven anomalous hall effect},\ }\href
  {https://doi.org/10.1103/PhysRevB.105.235142} {\bibfield  {journal} {\bibinfo
   {journal} {Phys. Rev. B}\ }\textbf {\bibinfo {volume} {105}},\ \bibinfo
  {pages} {235142} (\bibinfo {year} {2022})}\BibitemShut {NoStop}%
\bibitem [{\citenamefont {Salemi}\ and\ \citenamefont
  {Oppeneer}(2022{\natexlab{a}})}]{Oppeneer-PhysRevMaterials.6.095001}%
  \BibitemOpen
  \bibfield  {author} {\bibinfo {author} {\bibfnamefont {L.}~\bibnamefont
  {Salemi}}\ and\ \bibinfo {author} {\bibfnamefont {P.~M.}\ \bibnamefont
  {Oppeneer}},\ }\bibfield  {title} {\bibinfo {title} {First-principles theory
  of intrinsic spin and orbital hall and nernst effects in metallic monoatomic
  crystals},\ }\href {https://doi.org/10.1103/PhysRevMaterials.6.095001}
  {\bibfield  {journal} {\bibinfo  {journal} {Phys. Rev. Mater.}\ }\textbf
  {\bibinfo {volume} {6}},\ \bibinfo {pages} {095001} (\bibinfo {year}
  {2022}{\natexlab{a}})}\BibitemShut {NoStop}%
\bibitem [{\citenamefont {Kontani}\ \emph {et~al.}(2009)\citenamefont
  {Kontani}, \citenamefont {Tanaka}, \citenamefont {Hirashima}, \citenamefont
  {Yamada},\ and\ \citenamefont {Inoue}}]{Kontani-PhysRevLett.102.016601}%
  \BibitemOpen
  \bibfield  {author} {\bibinfo {author} {\bibfnamefont {H.}~\bibnamefont
  {Kontani}}, \bibinfo {author} {\bibfnamefont {T.}~\bibnamefont {Tanaka}},
  \bibinfo {author} {\bibfnamefont {D.~S.}\ \bibnamefont {Hirashima}}, \bibinfo
  {author} {\bibfnamefont {K.}~\bibnamefont {Yamada}},\ and\ \bibinfo {author}
  {\bibfnamefont {J.}~\bibnamefont {Inoue}},\ }\bibfield  {title} {\bibinfo
  {title} {Giant orbital hall effect in transition metals: Origin of large spin
  and anomalous hall effects},\ }\href
  {https://doi.org/10.1103/PhysRevLett.102.016601} {\bibfield  {journal}
  {\bibinfo  {journal} {Phys. Rev. Lett.}\ }\textbf {\bibinfo {volume} {102}},\
  \bibinfo {pages} {016601} (\bibinfo {year} {2009})}\BibitemShut {NoStop}%
\bibitem [{\citenamefont {Manchon}\ \emph {et~al.}(2023)\citenamefont
  {Manchon}, \citenamefont {Pezo}, \citenamefont {Kim},\ and\ \citenamefont
  {Lee}}]{manchon2023orbitalDiffusion}%
  \BibitemOpen
  \bibfield  {author} {\bibinfo {author} {\bibfnamefont {A.}~\bibnamefont
  {Manchon}}, \bibinfo {author} {\bibfnamefont {A.}~\bibnamefont {Pezo}},
  \bibinfo {author} {\bibfnamefont {K.-W.}\ \bibnamefont {Kim}},\ and\ \bibinfo
  {author} {\bibfnamefont {K.-J.}\ \bibnamefont {Lee}},\ }\href@noop {}
  {\bibinfo {title} {Orbital diffusion, polarization and swapping in
  centrosymmetric metals}} (\bibinfo {year} {2023}),\ \Eprint
  {https://arxiv.org/abs/2310.04763} {arXiv:2310.04763 [cond-mat.mes-hall]}
  \BibitemShut {NoStop}%
\bibitem [{\citenamefont {Mu}\ \emph {et~al.}(2021)\citenamefont {Mu},
  \citenamefont {Pan},\ and\ \citenamefont
  {Zhou}}]{Photoinduced-OrbitalHall-Mu2021}%
  \BibitemOpen
  \bibfield  {author} {\bibinfo {author} {\bibfnamefont {X.}~\bibnamefont
  {Mu}}, \bibinfo {author} {\bibfnamefont {Y.}~\bibnamefont {Pan}},\ and\
  \bibinfo {author} {\bibfnamefont {J.}~\bibnamefont {Zhou}},\ }\bibfield
  {title} {\bibinfo {title} {Pure bulk orbital and spin photocurrent in
  two-dimensional ferroelectric materials},\ }\bibfield  {journal} {\bibinfo
  {journal} {npj Computational Materials}\ }\textbf {\bibinfo {volume} {7}},\
  \href {https://doi.org/10.1038/s41524-021-00531-7}
  {10.1038/s41524-021-00531-7} (\bibinfo {year} {2021})\BibitemShut {NoStop}%
\bibitem [{\citenamefont {Santos}\ \emph {et~al.}(2024)\citenamefont {Santos},
  \citenamefont {Abr\~ao}, \citenamefont {Vieira}, \citenamefont {Mendes},
  \citenamefont {Rodr\'{\i}guez-Su\'arez},\ and\ \citenamefont
  {Azevedo}}]{santos2023exploring}%
  \BibitemOpen
  \bibfield  {author} {\bibinfo {author} {\bibfnamefont {E.}~\bibnamefont
  {Santos}}, \bibinfo {author} {\bibfnamefont {J.~E.}\ \bibnamefont {Abr\~ao}},
  \bibinfo {author} {\bibfnamefont {A.~S.}\ \bibnamefont {Vieira}}, \bibinfo
  {author} {\bibfnamefont {J.~B.~S.}\ \bibnamefont {Mendes}}, \bibinfo {author}
  {\bibfnamefont {R.~L.}\ \bibnamefont {Rodr\'{\i}guez-Su\'arez}},\ and\
  \bibinfo {author} {\bibfnamefont {A.}~\bibnamefont {Azevedo}},\ }\bibfield
  {title} {\bibinfo {title} {Exploring orbital-charge conversion mediated by
  interfaces with $\mathrm{Cu}{\mathrm{o}}_{x}$ through spin-orbital pumping},\
  }\href {https://doi.org/10.1103/PhysRevB.109.014420} {\bibfield  {journal}
  {\bibinfo  {journal} {Phys. Rev. B}\ }\textbf {\bibinfo {volume} {109}},\
  \bibinfo {pages} {014420} (\bibinfo {year} {2024})}\BibitemShut {NoStop}%
\bibitem [{\citenamefont {Han}\ \emph {et~al.}(2022)\citenamefont {Han},
  \citenamefont {Lee},\ and\ \citenamefont
  {Kim}}]{HWLee-PhysRevLett.128.176601}%
  \BibitemOpen
  \bibfield  {author} {\bibinfo {author} {\bibfnamefont {S.}~\bibnamefont
  {Han}}, \bibinfo {author} {\bibfnamefont {H.-W.}\ \bibnamefont {Lee}},\ and\
  \bibinfo {author} {\bibfnamefont {K.-W.}\ \bibnamefont {Kim}},\ }\bibfield
  {title} {\bibinfo {title} {Orbital dynamics in centrosymmetric systems},\
  }\href {https://doi.org/10.1103/PhysRevLett.128.176601} {\bibfield  {journal}
  {\bibinfo  {journal} {Phys. Rev. Lett.}\ }\textbf {\bibinfo {volume} {128}},\
  \bibinfo {pages} {176601} (\bibinfo {year} {2022})}\BibitemShut {NoStop}%
\bibitem [{\citenamefont {Shinada}\ and\ \citenamefont
  {Peters}(2023)}]{Orbital-PhysRevB.107.214109}%
  \BibitemOpen
  \bibfield  {author} {\bibinfo {author} {\bibfnamefont {K.}~\bibnamefont
  {Shinada}}\ and\ \bibinfo {author} {\bibfnamefont {R.}~\bibnamefont
  {Peters}},\ }\bibfield  {title} {\bibinfo {title} {Orbital
  gravitomagnetoelectric response and orbital magnetic quadrupole moment
  correction},\ }\href {https://doi.org/10.1103/PhysRevB.107.214109} {\bibfield
   {journal} {\bibinfo  {journal} {Phys. Rev. B}\ }\textbf {\bibinfo {volume}
  {107}},\ \bibinfo {pages} {214109} (\bibinfo {year} {2023})}\BibitemShut
  {NoStop}%
\bibitem [{\citenamefont {Leiva-Montecinos}\ \emph {et~al.}(2023)\citenamefont
  {Leiva-Montecinos}, \citenamefont {Henk}, \citenamefont {Mertig},\ and\
  \citenamefont {Johansson}}]{PhysRevResearch.5.043294}%
  \BibitemOpen
  \bibfield  {author} {\bibinfo {author} {\bibfnamefont {S.}~\bibnamefont
  {Leiva-Montecinos}}, \bibinfo {author} {\bibfnamefont {J.}~\bibnamefont
  {Henk}}, \bibinfo {author} {\bibfnamefont {I.}~\bibnamefont {Mertig}},\ and\
  \bibinfo {author} {\bibfnamefont {A.}~\bibnamefont {Johansson}},\ }\bibfield
  {title} {\bibinfo {title} {Spin and orbital edelstein effect in a bilayer
  system with rashba interaction},\ }\href
  {https://doi.org/10.1103/PhysRevResearch.5.043294} {\bibfield  {journal}
  {\bibinfo  {journal} {Phys. Rev. Res.}\ }\textbf {\bibinfo {volume} {5}},\
  \bibinfo {pages} {043294} (\bibinfo {year} {2023})}\BibitemShut {NoStop}%
\bibitem [{\citenamefont {Hayami}\ \emph {et~al.}(2018)\citenamefont {Hayami},
  \citenamefont {Yatsushiro}, \citenamefont {Yanagi},\ and\ \citenamefont
  {Kusunose}}]{Hayami-PhysRevB.98.165110}%
  \BibitemOpen
  \bibfield  {author} {\bibinfo {author} {\bibfnamefont {S.}~\bibnamefont
  {Hayami}}, \bibinfo {author} {\bibfnamefont {M.}~\bibnamefont {Yatsushiro}},
  \bibinfo {author} {\bibfnamefont {Y.}~\bibnamefont {Yanagi}},\ and\ \bibinfo
  {author} {\bibfnamefont {H.}~\bibnamefont {Kusunose}},\ }\bibfield  {title}
  {\bibinfo {title} {Classification of atomic-scale multipoles under
  crystallographic point groups and application to linear response tensors},\
  }\href {https://doi.org/10.1103/PhysRevB.98.165110} {\bibfield  {journal}
  {\bibinfo  {journal} {Phys. Rev. B}\ }\textbf {\bibinfo {volume} {98}},\
  \bibinfo {pages} {165110} (\bibinfo {year} {2018})}\BibitemShut {NoStop}%
\bibitem [{\citenamefont {Shinada}\ \emph {et~al.}(2023)\citenamefont
  {Shinada}, \citenamefont {Kofuji},\ and\ \citenamefont
  {Peters}}]{Peters-PhysRevB.107.094106}%
  \BibitemOpen
  \bibfield  {author} {\bibinfo {author} {\bibfnamefont {K.}~\bibnamefont
  {Shinada}}, \bibinfo {author} {\bibfnamefont {A.}~\bibnamefont {Kofuji}},\
  and\ \bibinfo {author} {\bibfnamefont {R.}~\bibnamefont {Peters}},\
  }\bibfield  {title} {\bibinfo {title} {Quantum theory of the intrinsic
  orbital magnetoelectric effect in itinerant electron systems at finite
  temperatures},\ }\href {https://doi.org/10.1103/PhysRevB.107.094106}
  {\bibfield  {journal} {\bibinfo  {journal} {Phys. Rev. B}\ }\textbf {\bibinfo
  {volume} {107}},\ \bibinfo {pages} {094106} (\bibinfo {year}
  {2023})}\BibitemShut {NoStop}%
\bibitem [{\citenamefont {Go}\ \emph {et~al.}(2023)\citenamefont {Go},
  \citenamefont {Ando}, \citenamefont {Pezo}, \citenamefont {Blügel},
  \citenamefont {Manchon},\ and\ \citenamefont
  {Mokrousov}}]{go2023orbital-Pumping}%
  \BibitemOpen
  \bibfield  {author} {\bibinfo {author} {\bibfnamefont {D.}~\bibnamefont
  {Go}}, \bibinfo {author} {\bibfnamefont {K.}~\bibnamefont {Ando}}, \bibinfo
  {author} {\bibfnamefont {A.}~\bibnamefont {Pezo}}, \bibinfo {author}
  {\bibfnamefont {S.}~\bibnamefont {Blügel}}, \bibinfo {author} {\bibfnamefont
  {A.}~\bibnamefont {Manchon}},\ and\ \bibinfo {author} {\bibfnamefont
  {Y.}~\bibnamefont {Mokrousov}},\ }\href@noop {} {\bibinfo {title} {Orbital
  pumping by magnetization dynamics in ferromagnets}} (\bibinfo {year}
  {2023}),\ \Eprint {https://arxiv.org/abs/2309.14817} {arXiv:2309.14817
  [cond-mat.mes-hall]} \BibitemShut {NoStop}%
\bibitem [{\citenamefont {Zeer}\ \emph {et~al.}(2024)\citenamefont {Zeer},
  \citenamefont {Go}, \citenamefont {Schmitz}, \citenamefont {Saunderson},
  \citenamefont {Wang}, \citenamefont {Ghabboun}, \citenamefont {Bl\"ugel},
  \citenamefont {Wulfhekel},\ and\ \citenamefont
  {Mokrousov}}]{Zeer-PhysRevResearch.6.013095}%
  \BibitemOpen
  \bibfield  {author} {\bibinfo {author} {\bibfnamefont {M.}~\bibnamefont
  {Zeer}}, \bibinfo {author} {\bibfnamefont {D.}~\bibnamefont {Go}}, \bibinfo
  {author} {\bibfnamefont {P.}~\bibnamefont {Schmitz}}, \bibinfo {author}
  {\bibfnamefont {T.~G.}\ \bibnamefont {Saunderson}}, \bibinfo {author}
  {\bibfnamefont {H.}~\bibnamefont {Wang}}, \bibinfo {author} {\bibfnamefont
  {J.}~\bibnamefont {Ghabboun}}, \bibinfo {author} {\bibfnamefont
  {S.}~\bibnamefont {Bl\"ugel}}, \bibinfo {author} {\bibfnamefont
  {W.}~\bibnamefont {Wulfhekel}},\ and\ \bibinfo {author} {\bibfnamefont
  {Y.}~\bibnamefont {Mokrousov}},\ }\bibfield  {title} {\bibinfo {title}
  {Promoting $p$-based hall effects by
  $p\text{\ensuremath{-}}d\text{\ensuremath{-}}f$ hybridization in gd-based
  dichalcogenides},\ }\href {https://doi.org/10.1103/PhysRevResearch.6.013095}
  {\bibfield  {journal} {\bibinfo  {journal} {Phys. Rev. Res.}\ }\textbf
  {\bibinfo {volume} {6}},\ \bibinfo {pages} {013095} (\bibinfo {year}
  {2024})}\BibitemShut {NoStop}%
\bibitem [{\citenamefont {Salemi}\ and\ \citenamefont
  {Oppeneer}(2022{\natexlab{b}})}]{Oppeneer-PhysRevB.106.024410}%
  \BibitemOpen
  \bibfield  {author} {\bibinfo {author} {\bibfnamefont {L.}~\bibnamefont
  {Salemi}}\ and\ \bibinfo {author} {\bibfnamefont {P.~M.}\ \bibnamefont
  {Oppeneer}},\ }\bibfield  {title} {\bibinfo {title} {Theory of magnetic spin
  and orbital hall and nernst effects in bulk ferromagnets},\ }\href
  {https://doi.org/10.1103/PhysRevB.106.024410} {\bibfield  {journal} {\bibinfo
   {journal} {Phys. Rev. B}\ }\textbf {\bibinfo {volume} {106}},\ \bibinfo
  {pages} {024410} (\bibinfo {year} {2022}{\natexlab{b}})}\BibitemShut
  {NoStop}%
\bibitem [{\citenamefont {Bhowal}\ and\ \citenamefont
  {Satpathy}(2020{\natexlab{a}})}]{Sayantika-PhysRevB.102.035409}%
  \BibitemOpen
  \bibfield  {author} {\bibinfo {author} {\bibfnamefont {S.}~\bibnamefont
  {Bhowal}}\ and\ \bibinfo {author} {\bibfnamefont {S.}~\bibnamefont
  {Satpathy}},\ }\bibfield  {title} {\bibinfo {title} {Intrinsic orbital and
  spin hall effects in monolayer transition metal dichalcogenides},\ }\href
  {https://doi.org/10.1103/PhysRevB.102.035409} {\bibfield  {journal} {\bibinfo
   {journal} {Phys. Rev. B}\ }\textbf {\bibinfo {volume} {102}},\ \bibinfo
  {pages} {035409} (\bibinfo {year} {2020}{\natexlab{a}})}\BibitemShut
  {NoStop}%
\bibitem [{\citenamefont {Bhowal}\ and\ \citenamefont
  {Satpathy}(2020{\natexlab{b}})}]{Sayantika-PhysRevB.101.121112}%
  \BibitemOpen
  \bibfield  {author} {\bibinfo {author} {\bibfnamefont {S.}~\bibnamefont
  {Bhowal}}\ and\ \bibinfo {author} {\bibfnamefont {S.}~\bibnamefont
  {Satpathy}},\ }\bibfield  {title} {\bibinfo {title} {Intrinsic orbital moment
  and prediction of a large orbital hall effect in two-dimensional transition
  metal dichalcogenides},\ }\href {https://doi.org/10.1103/PhysRevB.101.121112}
  {\bibfield  {journal} {\bibinfo  {journal} {Phys. Rev. B}\ }\textbf {\bibinfo
  {volume} {101}},\ \bibinfo {pages} {121112} (\bibinfo {year}
  {2020}{\natexlab{b}})}\BibitemShut {NoStop}%
\bibitem [{\citenamefont {Cysne}\ \emph
  {et~al.}(2021{\natexlab{a}})\citenamefont {Cysne}, \citenamefont
  {Guimar\~aes}, \citenamefont {Canonico}, \citenamefont {Rappoport},\ and\
  \citenamefont {Muniz}}]{Us-NRpband-PhysRevB.104.165403}%
  \BibitemOpen
  \bibfield  {author} {\bibinfo {author} {\bibfnamefont {T.~P.}\ \bibnamefont
  {Cysne}}, \bibinfo {author} {\bibfnamefont {F.~S.~M.}\ \bibnamefont
  {Guimar\~aes}}, \bibinfo {author} {\bibfnamefont {L.~M.}\ \bibnamefont
  {Canonico}}, \bibinfo {author} {\bibfnamefont {T.~G.}\ \bibnamefont
  {Rappoport}},\ and\ \bibinfo {author} {\bibfnamefont {R.~B.}\ \bibnamefont
  {Muniz}},\ }\bibfield  {title} {\bibinfo {title} {Orbital magnetoelectric
  effect in zigzag nanoribbons of $p$-band systems},\ }\href
  {https://doi.org/10.1103/PhysRevB.104.165403} {\bibfield  {journal} {\bibinfo
   {journal} {Phys. Rev. B}\ }\textbf {\bibinfo {volume} {104}},\ \bibinfo
  {pages} {165403} (\bibinfo {year} {2021}{\natexlab{a}})}\BibitemShut
  {NoStop}%
\bibitem [{\citenamefont {Cysne}\ \emph
  {et~al.}(2023{\natexlab{a}})\citenamefont {Cysne}, \citenamefont
  {Guimar\~aes}, \citenamefont {Canonico}, \citenamefont {Costa}, \citenamefont
  {Rappoport},\ and\ \citenamefont {Muniz}}]{Us-NR-TMD-PhysRevB.107.115402}%
  \BibitemOpen
  \bibfield  {author} {\bibinfo {author} {\bibfnamefont {T.~P.}\ \bibnamefont
  {Cysne}}, \bibinfo {author} {\bibfnamefont {F.~S.~M.}\ \bibnamefont
  {Guimar\~aes}}, \bibinfo {author} {\bibfnamefont {L.~M.}\ \bibnamefont
  {Canonico}}, \bibinfo {author} {\bibfnamefont {M.}~\bibnamefont {Costa}},
  \bibinfo {author} {\bibfnamefont {T.~G.}\ \bibnamefont {Rappoport}},\ and\
  \bibinfo {author} {\bibfnamefont {R.~B.}\ \bibnamefont {Muniz}},\ }\bibfield
  {title} {\bibinfo {title} {Orbital magnetoelectric effect in nanoribbons of
  transition metal dichalcogenides},\ }\href
  {https://doi.org/10.1103/PhysRevB.107.115402} {\bibfield  {journal} {\bibinfo
   {journal} {Phys. Rev. B}\ }\textbf {\bibinfo {volume} {107}},\ \bibinfo
  {pages} {115402} (\bibinfo {year} {2023}{\natexlab{a}})}\BibitemShut
  {NoStop}%
\bibitem [{\citenamefont {Zeer}\ \emph {et~al.}(2022)\citenamefont {Zeer},
  \citenamefont {Go}, \citenamefont {Carbone}, \citenamefont {Saunderson},
  \citenamefont {Redies}, \citenamefont {Kl\"aui}, \citenamefont {Ghabboun},
  \citenamefont {Wulfhekel}, \citenamefont {Bl\"ugel},\ and\ \citenamefont
  {Mokrousov}}]{Go-Mokrousov-PhysRevMaterials.6.074004}%
  \BibitemOpen
  \bibfield  {author} {\bibinfo {author} {\bibfnamefont {M.}~\bibnamefont
  {Zeer}}, \bibinfo {author} {\bibfnamefont {D.}~\bibnamefont {Go}}, \bibinfo
  {author} {\bibfnamefont {J.~P.}\ \bibnamefont {Carbone}}, \bibinfo {author}
  {\bibfnamefont {T.~G.}\ \bibnamefont {Saunderson}}, \bibinfo {author}
  {\bibfnamefont {M.}~\bibnamefont {Redies}}, \bibinfo {author} {\bibfnamefont
  {M.}~\bibnamefont {Kl\"aui}}, \bibinfo {author} {\bibfnamefont
  {J.}~\bibnamefont {Ghabboun}}, \bibinfo {author} {\bibfnamefont
  {W.}~\bibnamefont {Wulfhekel}}, \bibinfo {author} {\bibfnamefont
  {S.}~\bibnamefont {Bl\"ugel}},\ and\ \bibinfo {author} {\bibfnamefont
  {Y.}~\bibnamefont {Mokrousov}},\ }\bibfield  {title} {\bibinfo {title} {Spin
  and orbital transport in rare-earth dichalcogenides: The case of
  ${\mathrm{eus}}_{2}$},\ }\href
  {https://doi.org/10.1103/PhysRevMaterials.6.074004} {\bibfield  {journal}
  {\bibinfo  {journal} {Phys. Rev. Mater.}\ }\textbf {\bibinfo {volume} {6}},\
  \bibinfo {pages} {074004} (\bibinfo {year} {2022})}\BibitemShut {NoStop}%
\bibitem [{\citenamefont {Ji}\ \emph {et~al.}(2023)\citenamefont {Ji},
  \citenamefont {Yao}, \citenamefont {Quan}, \citenamefont {Wang},
  \citenamefont {Yang},\ and\ \citenamefont
  {Li}}]{Orbital-Chern-PhysRevB.108.224422}%
  \BibitemOpen
  \bibfield  {author} {\bibinfo {author} {\bibfnamefont {S.}~\bibnamefont
  {Ji}}, \bibinfo {author} {\bibfnamefont {R.}~\bibnamefont {Yao}}, \bibinfo
  {author} {\bibfnamefont {C.}~\bibnamefont {Quan}}, \bibinfo {author}
  {\bibfnamefont {Y.}~\bibnamefont {Wang}}, \bibinfo {author} {\bibfnamefont
  {J.}~\bibnamefont {Yang}},\ and\ \bibinfo {author} {\bibfnamefont
  {X.}~\bibnamefont {Li}},\ }\bibfield  {title} {\bibinfo {title} {Observing
  topological phase transition in ferromagnetic transition metal
  dichalcogenides},\ }\href {https://doi.org/10.1103/PhysRevB.108.224422}
  {\bibfield  {journal} {\bibinfo  {journal} {Phys. Rev. B}\ }\textbf {\bibinfo
  {volume} {108}},\ \bibinfo {pages} {224422} (\bibinfo {year}
  {2023})}\BibitemShut {NoStop}%
\bibitem [{\citenamefont {Ji}\ \emph {et~al.}(2024)\citenamefont {Ji},
  \citenamefont {Quan}, \citenamefont {Yao}, \citenamefont {Yang},\ and\
  \citenamefont {Li}}]{ji2023reversal}%
  \BibitemOpen
  \bibfield  {author} {\bibinfo {author} {\bibfnamefont {S.}~\bibnamefont
  {Ji}}, \bibinfo {author} {\bibfnamefont {C.}~\bibnamefont {Quan}}, \bibinfo
  {author} {\bibfnamefont {R.}~\bibnamefont {Yao}}, \bibinfo {author}
  {\bibfnamefont {J.}~\bibnamefont {Yang}},\ and\ \bibinfo {author}
  {\bibfnamefont {X.}~\bibnamefont {Li}},\ }\bibfield  {title} {\bibinfo
  {title} {Reversal of orbital hall conductivity and emergence of tunable
  topological quantum states in orbital hall insulators},\ }\href
  {https://doi.org/10.1103/PhysRevB.109.155407} {\bibfield  {journal} {\bibinfo
   {journal} {Phys. Rev. B}\ }\textbf {\bibinfo {volume} {109}},\ \bibinfo
  {pages} {155407} (\bibinfo {year} {2024})}\BibitemShut {NoStop}%
\bibitem [{\citenamefont {Xue}\ \emph {et~al.}(2020)\citenamefont {Xue},
  \citenamefont {Amin},\ and\ \citenamefont
  {Haney}}]{VHE_X_OHE-PhysRevB.102.161103}%
  \BibitemOpen
  \bibfield  {author} {\bibinfo {author} {\bibfnamefont {F.}~\bibnamefont
  {Xue}}, \bibinfo {author} {\bibfnamefont {V.}~\bibnamefont {Amin}},\ and\
  \bibinfo {author} {\bibfnamefont {P.~M.}\ \bibnamefont {Haney}},\ }\bibfield
  {title} {\bibinfo {title} {Imaging the valley and orbital hall effect in
  monolayer ${\mathrm{mos}}_{2}$},\ }\href
  {https://doi.org/10.1103/PhysRevB.102.161103} {\bibfield  {journal} {\bibinfo
   {journal} {Phys. Rev. B}\ }\textbf {\bibinfo {volume} {102}},\ \bibinfo
  {pages} {161103} (\bibinfo {year} {2020})}\BibitemShut {NoStop}%
\bibitem [{\citenamefont {Hayami}\ \emph {et~al.}(2016)\citenamefont {Hayami},
  \citenamefont {Kusunose},\ and\ \citenamefont
  {Motome}}]{Hayami_JPCMSymmetries_2016}%
  \BibitemOpen
  \bibfield  {author} {\bibinfo {author} {\bibfnamefont {S.}~\bibnamefont
  {Hayami}}, \bibinfo {author} {\bibfnamefont {H.}~\bibnamefont {Kusunose}},\
  and\ \bibinfo {author} {\bibfnamefont {Y.}~\bibnamefont {Motome}},\
  }\bibfield  {title} {\bibinfo {title} {Emergent spin-valley-orbital physics
  by spontaneous parity breaking},\ }\href
  {https://doi.org/10.1088/0953-8984/28/39/395601} {\bibfield  {journal}
  {\bibinfo  {journal} {Journal of Physics: Condensed Matter}\ }\textbf
  {\bibinfo {volume} {28}},\ \bibinfo {pages} {395601} (\bibinfo {year}
  {2016})}\BibitemShut {NoStop}%
\bibitem [{\citenamefont {Fonseca}\ \emph {et~al.}(2023)\citenamefont
  {Fonseca}, \citenamefont {Pereira},\ and\ \citenamefont
  {Barbosa}}]{AndersonBarbosa-PhysRevB.108.245105}%
  \BibitemOpen
  \bibfield  {author} {\bibinfo {author} {\bibfnamefont {D.~B.}\ \bibnamefont
  {Fonseca}}, \bibinfo {author} {\bibfnamefont {L.~L.~A.}\ \bibnamefont
  {Pereira}},\ and\ \bibinfo {author} {\bibfnamefont {A.~L.~R.}\ \bibnamefont
  {Barbosa}},\ }\bibfield  {title} {\bibinfo {title} {Orbital hall effect in
  mesoscopic devices},\ }\href {https://doi.org/10.1103/PhysRevB.108.245105}
  {\bibfield  {journal} {\bibinfo  {journal} {Phys. Rev. B}\ }\textbf {\bibinfo
  {volume} {108}},\ \bibinfo {pages} {245105} (\bibinfo {year}
  {2023})}\BibitemShut {NoStop}%
\bibitem [{\citenamefont {Schaefer}\ and\ \citenamefont
  {Nowack}(2021{\natexlab{a}})}]{OME-PhysRevB.103.224426}%
  \BibitemOpen
  \bibfield  {author} {\bibinfo {author} {\bibfnamefont {B.~T.}\ \bibnamefont
  {Schaefer}}\ and\ \bibinfo {author} {\bibfnamefont {K.~C.}\ \bibnamefont
  {Nowack}},\ }\bibfield  {title} {\bibinfo {title} {Electrically tunable and
  reversible magnetoelectric coupling in strained bilayer graphene},\ }\href
  {https://doi.org/10.1103/PhysRevB.103.224426} {\bibfield  {journal} {\bibinfo
   {journal} {Phys. Rev. B}\ }\textbf {\bibinfo {volume} {103}},\ \bibinfo
  {pages} {224426} (\bibinfo {year} {2021}{\natexlab{a}})}\BibitemShut
  {NoStop}%
\bibitem [{\citenamefont {Ghosh}\ and\ \citenamefont
  {Chittari}(2023)}]{ghosh2023orbital}%
  \BibitemOpen
  \bibfield  {author} {\bibinfo {author} {\bibfnamefont {S.}~\bibnamefont
  {Ghosh}}\ and\ \bibinfo {author} {\bibfnamefont {B.~L.}\ \bibnamefont
  {Chittari}},\ }\href@noop {} {\bibinfo {title} {Orbital hall conductivity in
  bilayer graphene}} (\bibinfo {year} {2023}),\ \Eprint
  {https://arxiv.org/abs/2311.06447} {arXiv:2311.06447 [cond-mat.mes-hall]}
  \BibitemShut {NoStop}%
\bibitem [{\citenamefont {Hayami}\ \emph {et~al.}(2014)\citenamefont {Hayami},
  \citenamefont {Kusunose},\ and\ \citenamefont
  {Motome}}]{Hayami_PhysRevB.90.081115}%
  \BibitemOpen
  \bibfield  {author} {\bibinfo {author} {\bibfnamefont {S.}~\bibnamefont
  {Hayami}}, \bibinfo {author} {\bibfnamefont {H.}~\bibnamefont {Kusunose}},\
  and\ \bibinfo {author} {\bibfnamefont {Y.}~\bibnamefont {Motome}},\
  }\bibfield  {title} {\bibinfo {title} {Spontaneous parity breaking in
  spin-orbital coupled systems},\ }\href
  {https://doi.org/10.1103/PhysRevB.90.081115} {\bibfield  {journal} {\bibinfo
  {journal} {Phys. Rev. B}\ }\textbf {\bibinfo {volume} {90}},\ \bibinfo
  {pages} {081115} (\bibinfo {year} {2014})}\BibitemShut {NoStop}%
\bibitem [{\citenamefont {Liu}\ and\ \citenamefont
  {Culcer}(2023)}]{liu2023dominance-Extrinsic}%
  \BibitemOpen
  \bibfield  {author} {\bibinfo {author} {\bibfnamefont {H.}~\bibnamefont
  {Liu}}\ and\ \bibinfo {author} {\bibfnamefont {D.}~\bibnamefont {Culcer}},\
  }\href@noop {} {\bibinfo {title} {Dominance of extrinsic scattering
  mechanisms in the orbital hall effect: graphene, transition metal
  dichalcogenides and topological antiferromagnets}} (\bibinfo {year} {2023}),\
  \Eprint {https://arxiv.org/abs/2308.14878} {arXiv:2308.14878
  [cond-mat.mes-hall]} \BibitemShut {NoStop}%
\bibitem [{\citenamefont {Pezo}\ \emph {et~al.}(2023)\citenamefont {Pezo},
  \citenamefont {Garc\'{\i}a~Ovalle},\ and\ \citenamefont
  {Manchon}}]{Manchon-PhysRevB.108.075427}%
  \BibitemOpen
  \bibfield  {author} {\bibinfo {author} {\bibfnamefont {A.}~\bibnamefont
  {Pezo}}, \bibinfo {author} {\bibfnamefont {D.}~\bibnamefont
  {Garc\'{\i}a~Ovalle}},\ and\ \bibinfo {author} {\bibfnamefont
  {A.}~\bibnamefont {Manchon}},\ }\bibfield  {title} {\bibinfo {title} {Orbital
  hall physics in two-dimensional dirac materials},\ }\href
  {https://doi.org/10.1103/PhysRevB.108.075427} {\bibfield  {journal} {\bibinfo
   {journal} {Phys. Rev. B}\ }\textbf {\bibinfo {volume} {108}},\ \bibinfo
  {pages} {075427} (\bibinfo {year} {2023})}\BibitemShut {NoStop}%
\bibitem [{\citenamefont {Sch\"uler}\ \emph {et~al.}(2022)\citenamefont
  {Sch\"uler}, \citenamefont {Pincelli}, \citenamefont {Dong}, \citenamefont
  {Devereaux}, \citenamefont {Wolf}, \citenamefont {Rettig}, \citenamefont
  {Ernstorfer},\ and\ \citenamefont {Beaulieu}}]{ARPES-PhysRevX.12.011019}%
  \BibitemOpen
  \bibfield  {author} {\bibinfo {author} {\bibfnamefont {M.}~\bibnamefont
  {Sch\"uler}}, \bibinfo {author} {\bibfnamefont {T.}~\bibnamefont {Pincelli}},
  \bibinfo {author} {\bibfnamefont {S.}~\bibnamefont {Dong}}, \bibinfo {author}
  {\bibfnamefont {T.~P.}\ \bibnamefont {Devereaux}}, \bibinfo {author}
  {\bibfnamefont {M.}~\bibnamefont {Wolf}}, \bibinfo {author} {\bibfnamefont
  {L.}~\bibnamefont {Rettig}}, \bibinfo {author} {\bibfnamefont
  {R.}~\bibnamefont {Ernstorfer}},\ and\ \bibinfo {author} {\bibfnamefont
  {S.}~\bibnamefont {Beaulieu}},\ }\bibfield  {title} {\bibinfo {title}
  {Polarization-modulated angle-resolved photoemission spectroscopy: Toward
  circular dichroism without circular photons and bloch wave-function
  reconstruction},\ }\href {https://doi.org/10.1103/PhysRevX.12.011019}
  {\bibfield  {journal} {\bibinfo  {journal} {Phys. Rev. X}\ }\textbf {\bibinfo
  {volume} {12}},\ \bibinfo {pages} {011019} (\bibinfo {year}
  {2022})}\BibitemShut {NoStop}%
\bibitem [{\citenamefont {Beaulieu}\ \emph {et~al.}(2020)\citenamefont
  {Beaulieu}, \citenamefont {Schusser}, \citenamefont {Dong}, \citenamefont
  {Sch\"uler}, \citenamefont {Pincelli}, \citenamefont {Dendzik}, \citenamefont
  {Maklar}, \citenamefont {Neef}, \citenamefont {Ebert}, \citenamefont
  {Hricovini}, \citenamefont {Wolf}, \citenamefont {Braun}, \citenamefont
  {Rettig}, \citenamefont {Min\'ar},\ and\ \citenamefont
  {Ernstorfer}}]{ARPES-PhysRevLett.125.216404}%
  \BibitemOpen
  \bibfield  {author} {\bibinfo {author} {\bibfnamefont {S.}~\bibnamefont
  {Beaulieu}}, \bibinfo {author} {\bibfnamefont {J.}~\bibnamefont {Schusser}},
  \bibinfo {author} {\bibfnamefont {S.}~\bibnamefont {Dong}}, \bibinfo {author}
  {\bibfnamefont {M.}~\bibnamefont {Sch\"uler}}, \bibinfo {author}
  {\bibfnamefont {T.}~\bibnamefont {Pincelli}}, \bibinfo {author}
  {\bibfnamefont {M.}~\bibnamefont {Dendzik}}, \bibinfo {author} {\bibfnamefont
  {J.}~\bibnamefont {Maklar}}, \bibinfo {author} {\bibfnamefont
  {A.}~\bibnamefont {Neef}}, \bibinfo {author} {\bibfnamefont {H.}~\bibnamefont
  {Ebert}}, \bibinfo {author} {\bibfnamefont {K.}~\bibnamefont {Hricovini}},
  \bibinfo {author} {\bibfnamefont {M.}~\bibnamefont {Wolf}}, \bibinfo {author}
  {\bibfnamefont {J.}~\bibnamefont {Braun}}, \bibinfo {author} {\bibfnamefont
  {L.}~\bibnamefont {Rettig}}, \bibinfo {author} {\bibfnamefont
  {J.}~\bibnamefont {Min\'ar}},\ and\ \bibinfo {author} {\bibfnamefont
  {R.}~\bibnamefont {Ernstorfer}},\ }\bibfield  {title} {\bibinfo {title}
  {Revealing hidden orbital pseudospin texture with time-reversal dichroism in
  photoelectron angular distributions},\ }\href
  {https://doi.org/10.1103/PhysRevLett.125.216404} {\bibfield  {journal}
  {\bibinfo  {journal} {Phys. Rev. Lett.}\ }\textbf {\bibinfo {volume} {125}},\
  \bibinfo {pages} {216404} (\bibinfo {year} {2020})}\BibitemShut {NoStop}%
\bibitem [{\citenamefont {Sch\"{u}ler}\ \emph {et~al.}(2023)\citenamefont
  {Sch\"{u}ler}, \citenamefont {Schmitt},\ and\ \citenamefont
  {Werner}}]{OAMText-Schler2023}%
  \BibitemOpen
  \bibfield  {author} {\bibinfo {author} {\bibfnamefont {M.}~\bibnamefont
  {Sch\"{u}ler}}, \bibinfo {author} {\bibfnamefont {T.}~\bibnamefont
  {Schmitt}},\ and\ \bibinfo {author} {\bibfnamefont {P.}~\bibnamefont
  {Werner}},\ }\bibfield  {title} {\bibinfo {title} {Probing magnetic orbitals
  and berry curvature with circular dichroism in resonant inelastic x-ray
  scattering},\ }\bibfield  {journal} {\bibinfo  {journal} {npj Quantum
  Materials}\ }\textbf {\bibinfo {volume} {8}},\ \href
  {https://doi.org/10.1038/s41535-023-00538-x} {10.1038/s41535-023-00538-x}
  (\bibinfo {year} {2023})\BibitemShut {NoStop}%
\bibitem [{\citenamefont {Canonico}\ \emph
  {et~al.}(2020{\natexlab{a}})\citenamefont {Canonico}, \citenamefont {Cysne},
  \citenamefont {Molina-Sanchez}, \citenamefont {Muniz},\ and\ \citenamefont
  {Rappoport}}]{Canonico-PhysRevB.101.161409}%
  \BibitemOpen
  \bibfield  {author} {\bibinfo {author} {\bibfnamefont {L.~M.}\ \bibnamefont
  {Canonico}}, \bibinfo {author} {\bibfnamefont {T.~P.}\ \bibnamefont {Cysne}},
  \bibinfo {author} {\bibfnamefont {A.}~\bibnamefont {Molina-Sanchez}},
  \bibinfo {author} {\bibfnamefont {R.~B.}\ \bibnamefont {Muniz}},\ and\
  \bibinfo {author} {\bibfnamefont {T.~G.}\ \bibnamefont {Rappoport}},\
  }\bibfield  {title} {\bibinfo {title} {Orbital hall insulating phase in
  transition metal dichalcogenide monolayers},\ }\href
  {https://doi.org/10.1103/PhysRevB.101.161409} {\bibfield  {journal} {\bibinfo
   {journal} {Phys. Rev. B}\ }\textbf {\bibinfo {volume} {101}},\ \bibinfo
  {pages} {161409} (\bibinfo {year} {2020}{\natexlab{a}})}\BibitemShut
  {NoStop}%
\bibitem [{\citenamefont {Canonico}\ \emph
  {et~al.}(2020{\natexlab{b}})\citenamefont {Canonico}, \citenamefont {Cysne},
  \citenamefont {Rappoport},\ and\ \citenamefont
  {Muniz}}]{Canonico-PhysRevB.101.075429}%
  \BibitemOpen
  \bibfield  {author} {\bibinfo {author} {\bibfnamefont {L.~M.}\ \bibnamefont
  {Canonico}}, \bibinfo {author} {\bibfnamefont {T.~P.}\ \bibnamefont {Cysne}},
  \bibinfo {author} {\bibfnamefont {T.~G.}\ \bibnamefont {Rappoport}},\ and\
  \bibinfo {author} {\bibfnamefont {R.~B.}\ \bibnamefont {Muniz}},\ }\bibfield
  {title} {\bibinfo {title} {Two-dimensional orbital hall insulators},\ }\href
  {https://doi.org/10.1103/PhysRevB.101.075429} {\bibfield  {journal} {\bibinfo
   {journal} {Phys. Rev. B}\ }\textbf {\bibinfo {volume} {101}},\ \bibinfo
  {pages} {075429} (\bibinfo {year} {2020}{\natexlab{b}})}\BibitemShut
  {NoStop}%
\bibitem [{\citenamefont {Cysne}\ \emph
  {et~al.}(2023{\natexlab{b}})\citenamefont {Cysne}, \citenamefont {Costa},
  \citenamefont {Nardelli}, \citenamefont {Muniz},\ and\ \citenamefont
  {Rappoport}}]{Us-BP-PhysRevB.108.165415}%
  \BibitemOpen
  \bibfield  {author} {\bibinfo {author} {\bibfnamefont {T.~P.}\ \bibnamefont
  {Cysne}}, \bibinfo {author} {\bibfnamefont {M.}~\bibnamefont {Costa}},
  \bibinfo {author} {\bibfnamefont {M.~B.}\ \bibnamefont {Nardelli}}, \bibinfo
  {author} {\bibfnamefont {R.~B.}\ \bibnamefont {Muniz}},\ and\ \bibinfo
  {author} {\bibfnamefont {T.~G.}\ \bibnamefont {Rappoport}},\ }\bibfield
  {title} {\bibinfo {title} {Ultrathin films of black phosphorus as suitable
  platforms for unambiguous observation of the orbital hall effect},\ }\href
  {https://doi.org/10.1103/PhysRevB.108.165415} {\bibfield  {journal} {\bibinfo
   {journal} {Phys. Rev. B}\ }\textbf {\bibinfo {volume} {108}},\ \bibinfo
  {pages} {165415} (\bibinfo {year} {2023}{\natexlab{b}})}\BibitemShut
  {NoStop}%
\bibitem [{\citenamefont {Cysne}\ \emph
  {et~al.}(2021{\natexlab{b}})\citenamefont {Cysne}, \citenamefont {Costa},
  \citenamefont {Canonico}, \citenamefont {Nardelli}, \citenamefont {Muniz},\
  and\ \citenamefont {Rappoport}}]{Cysne2021-PhysRevLett.126.056601}%
  \BibitemOpen
  \bibfield  {author} {\bibinfo {author} {\bibfnamefont {T.~P.}\ \bibnamefont
  {Cysne}}, \bibinfo {author} {\bibfnamefont {M.}~\bibnamefont {Costa}},
  \bibinfo {author} {\bibfnamefont {L.~M.}\ \bibnamefont {Canonico}}, \bibinfo
  {author} {\bibfnamefont {M.~B.}\ \bibnamefont {Nardelli}}, \bibinfo {author}
  {\bibfnamefont {R.~B.}\ \bibnamefont {Muniz}},\ and\ \bibinfo {author}
  {\bibfnamefont {T.~G.}\ \bibnamefont {Rappoport}},\ }\bibfield  {title}
  {\bibinfo {title} {Disentangling orbital and valley hall effects in bilayers
  of transition metal dichalcogenides},\ }\href
  {https://doi.org/10.1103/PhysRevLett.126.056601} {\bibfield  {journal}
  {\bibinfo  {journal} {Phys. Rev. Lett.}\ }\textbf {\bibinfo {volume} {126}},\
  \bibinfo {pages} {056601} (\bibinfo {year} {2021}{\natexlab{b}})}\BibitemShut
  {NoStop}%
\bibitem [{\citenamefont {Cysne}\ \emph {et~al.}(2022)\citenamefont {Cysne},
  \citenamefont {Bhowal}, \citenamefont {Vignale},\ and\ \citenamefont
  {Rappoport}}]{Cysne-Bhowal-Vignale-Rappoport-PhysRevB.105.195421}%
  \BibitemOpen
  \bibfield  {author} {\bibinfo {author} {\bibfnamefont {T.~P.}\ \bibnamefont
  {Cysne}}, \bibinfo {author} {\bibfnamefont {S.}~\bibnamefont {Bhowal}},
  \bibinfo {author} {\bibfnamefont {G.}~\bibnamefont {Vignale}},\ and\ \bibinfo
  {author} {\bibfnamefont {T.~G.}\ \bibnamefont {Rappoport}},\ }\bibfield
  {title} {\bibinfo {title} {Orbital hall effect in bilayer transition metal
  dichalcogenides: From the intra-atomic approximation to the bloch states
  orbital magnetic moment approach},\ }\href
  {https://doi.org/10.1103/PhysRevB.105.195421} {\bibfield  {journal} {\bibinfo
   {journal} {Phys. Rev. B}\ }\textbf {\bibinfo {volume} {105}},\ \bibinfo
  {pages} {195421} (\bibinfo {year} {2022})}\BibitemShut {NoStop}%
\bibitem [{\citenamefont {Qian}\ \emph {et~al.}(2022)\citenamefont {Qian},
  \citenamefont {Liu}, \citenamefont {Liu},\ and\ \citenamefont
  {Yao}}]{HOTI-TMD-PhysRevB.105.045417}%
  \BibitemOpen
  \bibfield  {author} {\bibinfo {author} {\bibfnamefont {S.}~\bibnamefont
  {Qian}}, \bibinfo {author} {\bibfnamefont {G.-B.}\ \bibnamefont {Liu}},
  \bibinfo {author} {\bibfnamefont {C.-C.}\ \bibnamefont {Liu}},\ and\ \bibinfo
  {author} {\bibfnamefont {Y.}~\bibnamefont {Yao}},\ }\bibfield  {title}
  {\bibinfo {title} {${C}_{n}$-symmetric higher-order topological crystalline
  insulators in atomically thin transition metal dichalcogenides},\ }\href
  {https://doi.org/10.1103/PhysRevB.105.045417} {\bibfield  {journal} {\bibinfo
   {journal} {Phys. Rev. B}\ }\textbf {\bibinfo {volume} {105}},\ \bibinfo
  {pages} {045417} (\bibinfo {year} {2022})}\BibitemShut {NoStop}%
\bibitem [{\citenamefont {Costa}\ \emph {et~al.}(2023)\citenamefont {Costa},
  \citenamefont {Focassio}, \citenamefont {Canonico}, \citenamefont {Cysne},
  \citenamefont {Schleder}, \citenamefont {Muniz}, \citenamefont {Fazzio},\
  and\ \citenamefont {Rappoport}}]{Us-HO-PRL-PhysRevLett.130.116204}%
  \BibitemOpen
  \bibfield  {author} {\bibinfo {author} {\bibfnamefont {M.}~\bibnamefont
  {Costa}}, \bibinfo {author} {\bibfnamefont {B.}~\bibnamefont {Focassio}},
  \bibinfo {author} {\bibfnamefont {L.~M.}\ \bibnamefont {Canonico}}, \bibinfo
  {author} {\bibfnamefont {T.~P.}\ \bibnamefont {Cysne}}, \bibinfo {author}
  {\bibfnamefont {G.~R.}\ \bibnamefont {Schleder}}, \bibinfo {author}
  {\bibfnamefont {R.~B.}\ \bibnamefont {Muniz}}, \bibinfo {author}
  {\bibfnamefont {A.}~\bibnamefont {Fazzio}},\ and\ \bibinfo {author}
  {\bibfnamefont {T.~G.}\ \bibnamefont {Rappoport}},\ }\bibfield  {title}
  {\bibinfo {title} {Connecting higher-order topology with the orbital hall
  effect in monolayers of transition metal dichalcogenides},\ }\href
  {https://doi.org/10.1103/PhysRevLett.130.116204} {\bibfield  {journal}
  {\bibinfo  {journal} {Phys. Rev. Lett.}\ }\textbf {\bibinfo {volume} {130}},\
  \bibinfo {pages} {116204} (\bibinfo {year} {2023})}\BibitemShut {NoStop}%
\bibitem [{\citenamefont {Gao}\ \emph {et~al.}(2020)\citenamefont {Gao},
  \citenamefont {Zhao}, \citenamefont {Alam~Ashik},\ and\ \citenamefont
  {Johnson}}]{Gao2020-ReviewBG}%
  \BibitemOpen
  \bibfield  {author} {\bibinfo {author} {\bibfnamefont {Z.}~\bibnamefont
  {Gao}}, \bibinfo {author} {\bibfnamefont {M.-Q.}\ \bibnamefont {Zhao}},
  \bibinfo {author} {\bibfnamefont {M.~M.}\ \bibnamefont {Alam~Ashik}},\ and\
  \bibinfo {author} {\bibfnamefont {A.~T.~C.}\ \bibnamefont {Johnson}},\
  }\bibfield  {title} {\bibinfo {title} {Recentadvances in the propertiesand
  synthesis of bilayer graphene and transition metal dichalcogenides},\ }\href
  {https://doi.org/10.1088/2515-7639/abb58d} {\bibfield  {journal} {\bibinfo
  {journal} {Journal of Physics: Materials}\ }\textbf {\bibinfo {volume} {3}},\
  \bibinfo {pages} {042003} (\bibinfo {year} {2020})}\BibitemShut {NoStop}%
\bibitem [{\citenamefont {Yan}(2015)}]{Yan2015-ReviewBG_Photonics}%
  \BibitemOpen
  \bibfield  {author} {\bibinfo {author} {\bibfnamefont {H.}~\bibnamefont
  {Yan}},\ }\bibfield  {title} {\bibinfo {title} {Bilayer graphene: physics and
  application outlook in photonics},\ }\href
  {https://doi.org/10.1515/nanoph-2014-0019} {\bibfield  {journal} {\bibinfo
  {journal} {Nanophotonics}\ }\textbf {\bibinfo {volume} {4}},\ \bibinfo
  {pages} {115–127} (\bibinfo {year} {2015})}\BibitemShut {NoStop}%
\bibitem [{\citenamefont {Yin}\ \emph {et~al.}(2022)\citenamefont {Yin},
  \citenamefont {Tan}, \citenamefont {Barcons-Ruiz}, \citenamefont {Torre},
  \citenamefont {Watanabe}, \citenamefont {Taniguchi}, \citenamefont {Song},
  \citenamefont {Hone},\ and\ \citenamefont {Koppens}}]{Koppens-Sience-2023}%
  \BibitemOpen
  \bibfield  {author} {\bibinfo {author} {\bibfnamefont {J.}~\bibnamefont
  {Yin}}, \bibinfo {author} {\bibfnamefont {C.}~\bibnamefont {Tan}}, \bibinfo
  {author} {\bibfnamefont {D.}~\bibnamefont {Barcons-Ruiz}}, \bibinfo {author}
  {\bibfnamefont {I.}~\bibnamefont {Torre}}, \bibinfo {author} {\bibfnamefont
  {K.}~\bibnamefont {Watanabe}}, \bibinfo {author} {\bibfnamefont
  {T.}~\bibnamefont {Taniguchi}}, \bibinfo {author} {\bibfnamefont {J.~C.~W.}\
  \bibnamefont {Song}}, \bibinfo {author} {\bibfnamefont {J.}~\bibnamefont
  {Hone}},\ and\ \bibinfo {author} {\bibfnamefont {F.~H.~L.}\ \bibnamefont
  {Koppens}},\ }\bibfield  {title} {\bibinfo {title} {Tunable and giant
  valley-selective hall effect in gapped bilayer graphene},\ }\href
  {https://doi.org/10.1126/science.abl4266} {\bibfield  {journal} {\bibinfo
  {journal} {Science}\ }\textbf {\bibinfo {volume} {375}},\ \bibinfo {pages}
  {1398} (\bibinfo {year} {2022})}\BibitemShut {NoStop}%
\bibitem [{\citenamefont {Song}\ and\ \citenamefont
  {Kats}(2016)}]{GiantHallPhotoconductivity-JSong_Nanolett}%
  \BibitemOpen
  \bibfield  {author} {\bibinfo {author} {\bibfnamefont {J.~C.~W.}\
  \bibnamefont {Song}}\ and\ \bibinfo {author} {\bibfnamefont {M.~A.}\
  \bibnamefont {Kats}},\ }\bibfield  {title} {\bibinfo {title} {Giant hall
  photoconductivity in narrow-gapped dirac materials},\ }\href
  {https://doi.org/10.1021/acs.nanolett.6b02559} {\bibfield  {journal}
  {\bibinfo  {journal} {Nano Letters}\ }\textbf {\bibinfo {volume} {16}},\
  \bibinfo {pages} {7346} (\bibinfo {year} {2016})},\ \bibinfo {note} {pMID:
  27960456},\ \Eprint
  {https://arxiv.org/abs/https://doi.org/10.1021/acs.nanolett.6b02559}
  {https://doi.org/10.1021/acs.nanolett.6b02559} \BibitemShut {NoStop}%
\bibitem [{\citenamefont {Souza}\ and\ \citenamefont
  {Vanderbilt}(2008)}]{ISouza-modTheo-PhysRevB.77.054438}%
  \BibitemOpen
  \bibfield  {author} {\bibinfo {author} {\bibfnamefont {I.}~\bibnamefont
  {Souza}}\ and\ \bibinfo {author} {\bibfnamefont {D.}~\bibnamefont
  {Vanderbilt}},\ }\bibfield  {title} {\bibinfo {title} {Dichroic $f$-sum rule
  and the orbital magnetization of crystals},\ }\href
  {https://doi.org/10.1103/PhysRevB.77.054438} {\bibfield  {journal} {\bibinfo
  {journal} {Phys. Rev. B}\ }\textbf {\bibinfo {volume} {77}},\ \bibinfo
  {pages} {054438} (\bibinfo {year} {2008})}\BibitemShut {NoStop}%
\bibitem [{\citenamefont {Xuan}\ and\ \citenamefont
  {Quek}(2020)}]{FXuan-modTheo-PhysRevResearch.2.033256}%
  \BibitemOpen
  \bibfield  {author} {\bibinfo {author} {\bibfnamefont {F.}~\bibnamefont
  {Xuan}}\ and\ \bibinfo {author} {\bibfnamefont {S.~Y.}\ \bibnamefont
  {Quek}},\ }\bibfield  {title} {\bibinfo {title} {Valley zeeman effect and
  landau levels in two-dimensional transition metal dichalcogenides},\ }\href
  {https://doi.org/10.1103/PhysRevResearch.2.033256} {\bibfield  {journal}
  {\bibinfo  {journal} {Phys. Rev. Res.}\ }\textbf {\bibinfo {volume} {2}},\
  \bibinfo {pages} {033256} (\bibinfo {year} {2020})}\BibitemShut {NoStop}%
\bibitem [{\citenamefont {Bhowal}\ and\ \citenamefont
  {Vignale}(2021)}]{Bhowal-Vignale-PhysRevB.103.195309}%
  \BibitemOpen
  \bibfield  {author} {\bibinfo {author} {\bibfnamefont {S.}~\bibnamefont
  {Bhowal}}\ and\ \bibinfo {author} {\bibfnamefont {G.}~\bibnamefont
  {Vignale}},\ }\bibfield  {title} {\bibinfo {title} {Orbital hall effect as an
  alternative to valley hall effect in gapped graphene},\ }\href
  {https://doi.org/10.1103/PhysRevB.103.195309} {\bibfield  {journal} {\bibinfo
   {journal} {Phys. Rev. B}\ }\textbf {\bibinfo {volume} {103}},\ \bibinfo
  {pages} {195309} (\bibinfo {year} {2021})}\BibitemShut {NoStop}%
\bibitem [{\citenamefont {Roche}\ \emph {et~al.}(2022)\citenamefont {Roche},
  \citenamefont {Power}, \citenamefont {Nikoli{\'{c}}}, \citenamefont
  {Garc{\'{\i}}a},\ and\ \citenamefont {Jauho}}]{Review-VHE-Roche2022}%
  \BibitemOpen
  \bibfield  {author} {\bibinfo {author} {\bibfnamefont {S.}~\bibnamefont
  {Roche}}, \bibinfo {author} {\bibfnamefont {S.~R.}\ \bibnamefont {Power}},
  \bibinfo {author} {\bibfnamefont {B.~K.}\ \bibnamefont {Nikoli{\'{c}}}},
  \bibinfo {author} {\bibfnamefont {J.~H.}\ \bibnamefont {Garc{\'{\i}}a}},\
  and\ \bibinfo {author} {\bibfnamefont {A.-P.}\ \bibnamefont {Jauho}},\
  }\bibfield  {title} {\bibinfo {title} {Have mysterious topological valley
  currents been observed in graphene superlattices?},\ }\href
  {https://doi.org/10.1088/2515-7639/ac452a} {\bibfield  {journal} {\bibinfo
  {journal} {Journal of Physics: Materials}\ }\textbf {\bibinfo {volume} {5}},\
  \bibinfo {pages} {021001} (\bibinfo {year} {2022})}\BibitemShut {NoStop}%
\bibitem [{\citenamefont {Wei}\ \emph {et~al.}(2014)\citenamefont {Wei},
  \citenamefont {Obstbaum}, \citenamefont {Ribow}, \citenamefont {Back},\ and\
  \citenamefont {Woltersdorf}}]{AC-SpincurrentDetection}%
  \BibitemOpen
  \bibfield  {author} {\bibinfo {author} {\bibfnamefont {D.}~\bibnamefont
  {Wei}}, \bibinfo {author} {\bibfnamefont {M.}~\bibnamefont {Obstbaum}},
  \bibinfo {author} {\bibfnamefont {M.}~\bibnamefont {Ribow}}, \bibinfo
  {author} {\bibfnamefont {C.~H.}\ \bibnamefont {Back}},\ and\ \bibinfo
  {author} {\bibfnamefont {G.}~\bibnamefont {Woltersdorf}},\ }\bibfield
  {title} {\bibinfo {title} {Spin hall voltages from a.c. and d.c. spin
  currents},\ }\bibfield  {journal} {\bibinfo  {journal} {Nature
  Communications}\ }\textbf {\bibinfo {volume} {5}},\ \href
  {https://doi.org/10.1038/ncomms4768} {10.1038/ncomms4768} (\bibinfo {year}
  {2014})\BibitemShut {NoStop}%
\bibitem [{\citenamefont {Weiler}\ \emph
  {et~al.}(2014{\natexlab{a}})\citenamefont {Weiler}, \citenamefont {Shaw},
  \citenamefont {Nembach},\ and\ \citenamefont
  {Silva}}]{AC-SHE-PhysRevLett.113.157204}%
  \BibitemOpen
  \bibfield  {author} {\bibinfo {author} {\bibfnamefont {M.}~\bibnamefont
  {Weiler}}, \bibinfo {author} {\bibfnamefont {J.~M.}\ \bibnamefont {Shaw}},
  \bibinfo {author} {\bibfnamefont {H.~T.}\ \bibnamefont {Nembach}},\ and\
  \bibinfo {author} {\bibfnamefont {T.~J.}\ \bibnamefont {Silva}},\ }\bibfield
  {title} {\bibinfo {title} {Phase-sensitive detection of spin pumping via the
  ac inverse spin hall effect},\ }\href
  {https://doi.org/10.1103/PhysRevLett.113.157204} {\bibfield  {journal}
  {\bibinfo  {journal} {Phys. Rev. Lett.}\ }\textbf {\bibinfo {volume} {113}},\
  \bibinfo {pages} {157204} (\bibinfo {year} {2014}{\natexlab{a}})}\BibitemShut
  {NoStop}%
\bibitem [{\citenamefont {Tabert}\ and\ \citenamefont
  {Nicol}(2013)}]{AC_SHE-PhysRevB.87.235426}%
  \BibitemOpen
  \bibfield  {author} {\bibinfo {author} {\bibfnamefont {C.~J.}\ \bibnamefont
  {Tabert}}\ and\ \bibinfo {author} {\bibfnamefont {E.~J.}\ \bibnamefont
  {Nicol}},\ }\bibfield  {title} {\bibinfo {title} {Ac/dc spin and valley hall
  effects in silicene and germanene},\ }\href
  {https://doi.org/10.1103/PhysRevB.87.235426} {\bibfield  {journal} {\bibinfo
  {journal} {Phys. Rev. B}\ }\textbf {\bibinfo {volume} {87}},\ \bibinfo
  {pages} {235426} (\bibinfo {year} {2013})}\BibitemShut {NoStop}%
\bibitem [{\citenamefont {Chen}(2023)}]{Chen-JPCM_ACSHE_2023}%
  \BibitemOpen
  \bibfield  {author} {\bibinfo {author} {\bibfnamefont {W.}~\bibnamefont
  {Chen}},\ }\bibfield  {title} {\bibinfo {title} {Optical absorption
  measurement of spin berry curvature and spin chern marker},\ }\href
  {https://doi.org/10.1088/1361-648X/acba72} {\bibfield  {journal} {\bibinfo
  {journal} {Journal of Physics: Condensed Matter}\ }\textbf {\bibinfo {volume}
  {35}},\ \bibinfo {pages} {155601} (\bibinfo {year} {2023})}\BibitemShut
  {NoStop}%
\bibitem [{\citenamefont {Guo}\ \emph {et~al.}(2005)\citenamefont {Guo},
  \citenamefont {Yao},\ and\ \citenamefont
  {Niu}}]{AC-SHC-PhysRevLett.94.226601}%
  \BibitemOpen
  \bibfield  {author} {\bibinfo {author} {\bibfnamefont {G.~Y.}\ \bibnamefont
  {Guo}}, \bibinfo {author} {\bibfnamefont {Y.}~\bibnamefont {Yao}},\ and\
  \bibinfo {author} {\bibfnamefont {Q.}~\bibnamefont {Niu}},\ }\bibfield
  {title} {\bibinfo {title} {Ab initio calculation of the intrinsic spin hall
  effect in semiconductors},\ }\href
  {https://doi.org/10.1103/PhysRevLett.94.226601} {\bibfield  {journal}
  {\bibinfo  {journal} {Phys. Rev. Lett.}\ }\textbf {\bibinfo {volume} {94}},\
  \bibinfo {pages} {226601} (\bibinfo {year} {2005})}\BibitemShut {NoStop}%
\bibitem [{\citenamefont {Guimar{\~{a}}es}\ \emph {et~al.}(2017)\citenamefont
  {Guimar{\~{a}}es}, \citenamefont {dos Santos~Dias}, \citenamefont {Bouaziz},
  \citenamefont {Costa}, \citenamefont {Muniz},\ and\ \citenamefont
  {Lounis}}]{Bechara-THz-2017}%
  \BibitemOpen
  \bibfield  {author} {\bibinfo {author} {\bibfnamefont {F.~S.~M.}\
  \bibnamefont {Guimar{\~{a}}es}}, \bibinfo {author} {\bibfnamefont
  {M.}~\bibnamefont {dos Santos~Dias}}, \bibinfo {author} {\bibfnamefont
  {J.}~\bibnamefont {Bouaziz}}, \bibinfo {author} {\bibfnamefont {A.~T.}\
  \bibnamefont {Costa}}, \bibinfo {author} {\bibfnamefont {R.~B.}\ \bibnamefont
  {Muniz}},\ and\ \bibinfo {author} {\bibfnamefont {S.}~\bibnamefont
  {Lounis}},\ }\bibfield  {title} {\bibinfo {title} {Dynamical amplification of
  magnetoresistances and hall currents up to the {THz} regime},\ }\bibfield
  {journal} {\bibinfo  {journal} {Scientific Reports}\ }\textbf {\bibinfo
  {volume} {7}},\ \href {https://doi.org/10.1038/s41598-017-03924-1}
  {10.1038/s41598-017-03924-1} (\bibinfo {year} {2017})\BibitemShut {NoStop}%
\bibitem [{\citenamefont {Habara}\ and\ \citenamefont
  {Wakabayashi}(2021)}]{AC_SHE-TMD_PhysRevB.103.L161410}%
  \BibitemOpen
  \bibfield  {author} {\bibinfo {author} {\bibfnamefont {R.}~\bibnamefont
  {Habara}}\ and\ \bibinfo {author} {\bibfnamefont {K.}~\bibnamefont
  {Wakabayashi}},\ }\bibfield  {title} {\bibinfo {title} {Optically induced
  spin current in monolayer ${\mathrm{nbse}}_{2}$},\ }\href
  {https://doi.org/10.1103/PhysRevB.103.L161410} {\bibfield  {journal}
  {\bibinfo  {journal} {Phys. Rev. B}\ }\textbf {\bibinfo {volume} {103}},\
  \bibinfo {pages} {L161410} (\bibinfo {year} {2021})}\BibitemShut {NoStop}%
\bibitem [{\citenamefont {Phong}\ \emph {et~al.}(2019)\citenamefont {Phong},
  \citenamefont {Addison}, \citenamefont {Ahn}, \citenamefont {Min},
  \citenamefont {Agarwal},\ and\ \citenamefont
  {Mele}}]{MeleOrbitronics-PhysRevLett.123.236403}%
  \BibitemOpen
  \bibfield  {author} {\bibinfo {author} {\bibfnamefont {V.~o.~T.}\
  \bibnamefont {Phong}}, \bibinfo {author} {\bibfnamefont {Z.}~\bibnamefont
  {Addison}}, \bibinfo {author} {\bibfnamefont {S.}~\bibnamefont {Ahn}},
  \bibinfo {author} {\bibfnamefont {H.}~\bibnamefont {Min}}, \bibinfo {author}
  {\bibfnamefont {R.}~\bibnamefont {Agarwal}},\ and\ \bibinfo {author}
  {\bibfnamefont {E.~J.}\ \bibnamefont {Mele}},\ }\bibfield  {title} {\bibinfo
  {title} {Optically controlled orbitronics on a triangular lattice},\ }\href
  {https://doi.org/10.1103/PhysRevLett.123.236403} {\bibfield  {journal}
  {\bibinfo  {journal} {Phys. Rev. Lett.}\ }\textbf {\bibinfo {volume} {123}},\
  \bibinfo {pages} {236403} (\bibinfo {year} {2019})}\BibitemShut {NoStop}%
\bibitem [{\citenamefont {Busch}\ \emph
  {et~al.}(2023{\natexlab{a}})\citenamefont {Busch}, \citenamefont
  {Ziolkowski}, \citenamefont {Mertig},\ and\ \citenamefont
  {Henk}}]{Mertig-PhysRevB.108.184401}%
  \BibitemOpen
  \bibfield  {author} {\bibinfo {author} {\bibfnamefont {O.}~\bibnamefont
  {Busch}}, \bibinfo {author} {\bibfnamefont {F.}~\bibnamefont {Ziolkowski}},
  \bibinfo {author} {\bibfnamefont {I.}~\bibnamefont {Mertig}},\ and\ \bibinfo
  {author} {\bibfnamefont {J.}~\bibnamefont {Henk}},\ }\bibfield  {title}
  {\bibinfo {title} {Ultrafast dynamics of electrons excited by femtosecond
  laser pulses: Spin polarization and spin-polarized currents},\ }\href
  {https://doi.org/10.1103/PhysRevB.108.184401} {\bibfield  {journal} {\bibinfo
   {journal} {Phys. Rev. B}\ }\textbf {\bibinfo {volume} {108}},\ \bibinfo
  {pages} {184401} (\bibinfo {year} {2023}{\natexlab{a}})}\BibitemShut
  {NoStop}%
\bibitem [{\citenamefont {Busch}\ \emph
  {et~al.}(2023{\natexlab{b}})\citenamefont {Busch}, \citenamefont
  {Ziolkowski}, \citenamefont {Mertig},\ and\ \citenamefont
  {Henk}}]{Mertig-PhysRevB.108.104408}%
  \BibitemOpen
  \bibfield  {author} {\bibinfo {author} {\bibfnamefont {O.}~\bibnamefont
  {Busch}}, \bibinfo {author} {\bibfnamefont {F.}~\bibnamefont {Ziolkowski}},
  \bibinfo {author} {\bibfnamefont {I.}~\bibnamefont {Mertig}},\ and\ \bibinfo
  {author} {\bibfnamefont {J.}~\bibnamefont {Henk}},\ }\bibfield  {title}
  {\bibinfo {title} {Ultrafast dynamics of orbital angular momentum of
  electrons induced by femtosecond laser pulses: Generation and transfer across
  interfaces},\ }\href {https://doi.org/10.1103/PhysRevB.108.104408} {\bibfield
   {journal} {\bibinfo  {journal} {Phys. Rev. B}\ }\textbf {\bibinfo {volume}
  {108}},\ \bibinfo {pages} {104408} (\bibinfo {year}
  {2023}{\natexlab{b}})}\BibitemShut {NoStop}%
\bibitem [{\citenamefont {Hamamera}\ \emph {et~al.}(2023)\citenamefont
  {Hamamera}, \citenamefont {Guimarães}, \citenamefont {dos Santos~Dias},\
  and\ \citenamefont {Lounis}}]{hamamera2023ultrafast}%
  \BibitemOpen
  \bibfield  {author} {\bibinfo {author} {\bibfnamefont {H.}~\bibnamefont
  {Hamamera}}, \bibinfo {author} {\bibfnamefont {F.~S.~M.}\ \bibnamefont
  {Guimarães}}, \bibinfo {author} {\bibfnamefont {M.}~\bibnamefont {dos
  Santos~Dias}},\ and\ \bibinfo {author} {\bibfnamefont {S.}~\bibnamefont
  {Lounis}},\ }\href@noop {} {\bibinfo {title} {Ultrafast light-induced
  magnetization in non-magnetic films: from orbital and spin hall phenomena to
  the inverse faraday effect}} (\bibinfo {year} {2023}),\ \Eprint
  {https://arxiv.org/abs/2312.07888} {arXiv:2312.07888 [cond-mat.mtrl-sci]}
  \BibitemShut {NoStop}%
\bibitem [{\citenamefont {Busch}\ \emph
  {et~al.}(2023{\natexlab{c}})\citenamefont {Busch}, \citenamefont
  {Ziolkowski}, \citenamefont {Göbel}, \citenamefont {Mertig},\ and\
  \citenamefont {Henk}}]{Ultrafast-OHE-Mertig}%
  \BibitemOpen
  \bibfield  {author} {\bibinfo {author} {\bibfnamefont {O.}~\bibnamefont
  {Busch}}, \bibinfo {author} {\bibfnamefont {F.}~\bibnamefont {Ziolkowski}},
  \bibinfo {author} {\bibfnamefont {B.}~\bibnamefont {Göbel}}, \bibinfo
  {author} {\bibfnamefont {I.}~\bibnamefont {Mertig}},\ and\ \bibinfo {author}
  {\bibfnamefont {J.}~\bibnamefont {Henk}},\ }\href@noop {} {\bibinfo {title}
  {Ultrafast orbital hall effect in metallic nanoribbons}} (\bibinfo {year}
  {2023}{\natexlab{c}}),\ \Eprint {https://arxiv.org/abs/2307.08444}
  {arXiv:2307.08444 [cond-mat.mtrl-sci]} \BibitemShut {NoStop}%
\bibitem [{\citenamefont {Xu}\ \emph {et~al.}(2023)\citenamefont {Xu},
  \citenamefont {Zhang}, \citenamefont {Fert}, \citenamefont {Jaffres},
  \citenamefont {Liu}, \citenamefont {Xu}, \citenamefont {Jiang}, \citenamefont
  {Cheng},\ and\ \citenamefont {Zhao}}]{Fert-xu2023orbitronics}%
  \BibitemOpen
  \bibfield  {author} {\bibinfo {author} {\bibfnamefont {Y.}~\bibnamefont
  {Xu}}, \bibinfo {author} {\bibfnamefont {F.}~\bibnamefont {Zhang}}, \bibinfo
  {author} {\bibfnamefont {A.}~\bibnamefont {Fert}}, \bibinfo {author}
  {\bibfnamefont {H.-Y.}\ \bibnamefont {Jaffres}}, \bibinfo {author}
  {\bibfnamefont {Y.}~\bibnamefont {Liu}}, \bibinfo {author} {\bibfnamefont
  {R.}~\bibnamefont {Xu}}, \bibinfo {author} {\bibfnamefont {Y.}~\bibnamefont
  {Jiang}}, \bibinfo {author} {\bibfnamefont {H.}~\bibnamefont {Cheng}},\ and\
  \bibinfo {author} {\bibfnamefont {W.}~\bibnamefont {Zhao}},\ }\href@noop {}
  {\bibinfo {title} {Orbitronics: Light-induced orbit currents in terahertz
  emission experiments}} (\bibinfo {year} {2023}),\ \Eprint
  {https://arxiv.org/abs/2307.03490} {arXiv:2307.03490 [cond-mat.mes-hall]}
  \BibitemShut {NoStop}%
\bibitem [{\citenamefont {Seifert}\ \emph {et~al.}(2023)\citenamefont
  {Seifert}, \citenamefont {Go}, \citenamefont {Hayashi}, \citenamefont
  {Rouzegar}, \citenamefont {Freimuth}, \citenamefont {Ando}, \citenamefont
  {Mokrousov},\ and\ \citenamefont
  {Kampfrath}}]{Go-Thz-emission-10.1038/s41565-023-01470-8}%
  \BibitemOpen
  \bibfield  {author} {\bibinfo {author} {\bibfnamefont {T.~S.}\ \bibnamefont
  {Seifert}}, \bibinfo {author} {\bibfnamefont {D.}~\bibnamefont {Go}},
  \bibinfo {author} {\bibfnamefont {H.}~\bibnamefont {Hayashi}}, \bibinfo
  {author} {\bibfnamefont {R.}~\bibnamefont {Rouzegar}}, \bibinfo {author}
  {\bibfnamefont {F.}~\bibnamefont {Freimuth}}, \bibinfo {author}
  {\bibfnamefont {K.}~\bibnamefont {Ando}}, \bibinfo {author} {\bibfnamefont
  {Y.}~\bibnamefont {Mokrousov}},\ and\ \bibinfo {author} {\bibfnamefont
  {T.}~\bibnamefont {Kampfrath}},\ }\bibfield  {title} {\bibinfo {title}
  {Time-domain observation of ballistic orbital-angular-momentum currents with
  giant relaxation length in tungsten},\ }\href
  {https://doi.org/10.1038/s41565-023-01470-8} {\bibfield  {journal} {\bibinfo
  {journal} {Nature Nanotechnology}\ }\textbf {\bibinfo {volume} {18}},\
  \bibinfo {pages} {1132–1138} (\bibinfo {year} {2023})}\BibitemShut
  {NoStop}%
\bibitem [{\citenamefont {Wang}\ \emph {et~al.}(2023)\citenamefont {Wang},
  \citenamefont {Feng}, \citenamefont {Yang}, \citenamefont {Zhang},
  \citenamefont {Liu}, \citenamefont {Xu}, \citenamefont {Jia}, \citenamefont
  {Wu}, \citenamefont {Yu}, \citenamefont {Xu},\ and\ \citenamefont
  {Jiang}}]{IOHE-THz-s41535-023-00559-6}%
  \BibitemOpen
  \bibfield  {author} {\bibinfo {author} {\bibfnamefont {P.}~\bibnamefont
  {Wang}}, \bibinfo {author} {\bibfnamefont {Z.}~\bibnamefont {Feng}}, \bibinfo
  {author} {\bibfnamefont {Y.}~\bibnamefont {Yang}}, \bibinfo {author}
  {\bibfnamefont {D.}~\bibnamefont {Zhang}}, \bibinfo {author} {\bibfnamefont
  {Q.}~\bibnamefont {Liu}}, \bibinfo {author} {\bibfnamefont {Z.}~\bibnamefont
  {Xu}}, \bibinfo {author} {\bibfnamefont {Z.}~\bibnamefont {Jia}}, \bibinfo
  {author} {\bibfnamefont {Y.}~\bibnamefont {Wu}}, \bibinfo {author}
  {\bibfnamefont {G.}~\bibnamefont {Yu}}, \bibinfo {author} {\bibfnamefont
  {X.}~\bibnamefont {Xu}},\ and\ \bibinfo {author} {\bibfnamefont
  {Y.}~\bibnamefont {Jiang}},\ }\bibfield  {title} {\bibinfo {title} {Inverse
  orbital hall effect and orbitronic terahertz emission observed in the
  materials with weak spin-orbit coupling},\ }\bibfield  {journal} {\bibinfo
  {journal} {npj Quantum Materials}\ }\textbf {\bibinfo {volume} {8}},\ \href
  {https://doi.org/10.1038/s41535-023-00559-6} {10.1038/s41535-023-00559-6}
  (\bibinfo {year} {2023})\BibitemShut {NoStop}%
\bibitem [{\citenamefont {Mishra}\ \emph {et~al.}(2024)\citenamefont {Mishra},
  \citenamefont {Lourembam}, \citenamefont {Lin},\ and\ \citenamefont
  {Singh}}]{mishra2024active}%
  \BibitemOpen
  \bibfield  {author} {\bibinfo {author} {\bibfnamefont {S.~S.}\ \bibnamefont
  {Mishra}}, \bibinfo {author} {\bibfnamefont {J.}~\bibnamefont {Lourembam}},
  \bibinfo {author} {\bibfnamefont {D.~J.~X.}\ \bibnamefont {Lin}},\ and\
  \bibinfo {author} {\bibfnamefont {R.}~\bibnamefont {Singh}},\ }\href@noop {}
  {\bibinfo {title} {Active control of ballistic orbital transport}} (\bibinfo
  {year} {2024}),\ \Eprint {https://arxiv.org/abs/2401.08373} {arXiv:2401.08373
  [cond-mat.mtrl-sci]} \BibitemShut {NoStop}%
\bibitem [{\citenamefont
  {McCann}(2006)}]{2L-graphene-McCann-PhysRevB.74.161403}%
  \BibitemOpen
  \bibfield  {author} {\bibinfo {author} {\bibfnamefont {E.}~\bibnamefont
  {McCann}},\ }\bibfield  {title} {\bibinfo {title} {Asymmetry gap in the
  electronic band structure of bilayer graphene},\ }\href
  {https://doi.org/10.1103/PhysRevB.74.161403} {\bibfield  {journal} {\bibinfo
  {journal} {Phys. Rev. B}\ }\textbf {\bibinfo {volume} {74}},\ \bibinfo
  {pages} {161403} (\bibinfo {year} {2006})}\BibitemShut {NoStop}%
\bibitem [{\citenamefont {Tang}\ and\ \citenamefont
  {Bauer}(2024)}]{tang2024role}%
  \BibitemOpen
  \bibfield  {author} {\bibinfo {author} {\bibfnamefont {P.}~\bibnamefont
  {Tang}}\ and\ \bibinfo {author} {\bibfnamefont {G.~E.~W.}\ \bibnamefont
  {Bauer}},\ }\href@noop {} {\bibinfo {title} {Role of disorder in the
  intrinsic orbital hall effect}} (\bibinfo {year} {2024}),\ \Eprint
  {https://arxiv.org/abs/2401.17620} {arXiv:2401.17620 [cond-mat.mes-hall]}
  \BibitemShut {NoStop}%
\bibitem [{\citenamefont {McCann}\ and\ \citenamefont
  {Koshino}(2013)}]{McCann_2013-Eletrostatic}%
  \BibitemOpen
  \bibfield  {author} {\bibinfo {author} {\bibfnamefont {E.}~\bibnamefont
  {McCann}}\ and\ \bibinfo {author} {\bibfnamefont {M.}~\bibnamefont
  {Koshino}},\ }\bibfield  {title} {\bibinfo {title} {The electronic properties
  of bilayer graphene},\ }\href {https://doi.org/10.1088/0034-4885/76/5/056503}
  {\bibfield  {journal} {\bibinfo  {journal} {Reports on Progress in Physics}\
  }\textbf {\bibinfo {volume} {76}},\ \bibinfo {pages} {056503} (\bibinfo
  {year} {2013})}\BibitemShut {NoStop}%
\bibitem [{\citenamefont {Schaefer}\ and\ \citenamefont
  {Nowack}(2021{\natexlab{b}})}]{Nowack-PhysRevB.103.224426}%
  \BibitemOpen
  \bibfield  {author} {\bibinfo {author} {\bibfnamefont {B.~T.}\ \bibnamefont
  {Schaefer}}\ and\ \bibinfo {author} {\bibfnamefont {K.~C.}\ \bibnamefont
  {Nowack}},\ }\bibfield  {title} {\bibinfo {title} {Electrically tunable and
  reversible magnetoelectric coupling in strained bilayer graphene},\ }\href
  {https://doi.org/10.1103/PhysRevB.103.224426} {\bibfield  {journal} {\bibinfo
   {journal} {Phys. Rev. B}\ }\textbf {\bibinfo {volume} {103}},\ \bibinfo
  {pages} {224426} (\bibinfo {year} {2021}{\natexlab{b}})}\BibitemShut
  {NoStop}%
\bibitem [{\citenamefont {Iwasaki}\ \emph {et~al.}(2022)\citenamefont
  {Iwasaki}, \citenamefont {Morita}, \citenamefont {Watanabe},\ and\
  \citenamefont {Taniguchi}}]{Exp-DualgateBLG-PhysRevB.106.165134}%
  \BibitemOpen
  \bibfield  {author} {\bibinfo {author} {\bibfnamefont {T.}~\bibnamefont
  {Iwasaki}}, \bibinfo {author} {\bibfnamefont {Y.}~\bibnamefont {Morita}},
  \bibinfo {author} {\bibfnamefont {K.}~\bibnamefont {Watanabe}},\ and\
  \bibinfo {author} {\bibfnamefont {T.}~\bibnamefont {Taniguchi}},\ }\bibfield
  {title} {\bibinfo {title} {Dual-gated hbn/bilayer-graphene superlattices and
  the transitions between the insulating phases at the charge neutrality
  point},\ }\href {https://doi.org/10.1103/PhysRevB.106.165134} {\bibfield
  {journal} {\bibinfo  {journal} {Phys. Rev. B}\ }\textbf {\bibinfo {volume}
  {106}},\ \bibinfo {pages} {165134} (\bibinfo {year} {2022})}\BibitemShut
  {NoStop}%
\bibitem [{\citenamefont {Kort-Kamp}(2017)}]{Wilton-PhysRevLett.119.147401}%
  \BibitemOpen
  \bibfield  {author} {\bibinfo {author} {\bibfnamefont {W.~J.~M.}\
  \bibnamefont {Kort-Kamp}},\ }\bibfield  {title} {\bibinfo {title}
  {Topological phase transitions in the photonic spin hall effect},\ }\href
  {https://doi.org/10.1103/PhysRevLett.119.147401} {\bibfield  {journal}
  {\bibinfo  {journal} {Phys. Rev. Lett.}\ }\textbf {\bibinfo {volume} {119}},\
  \bibinfo {pages} {147401} (\bibinfo {year} {2017})}\BibitemShut {NoStop}%
\bibitem [{\citenamefont {Salvador-Sánchez}\ \emph {et~al.}(2022)\citenamefont
  {Salvador-Sánchez}, \citenamefont {Canonico}, \citenamefont
  {Pérez-Rodríguez}, \citenamefont {Cysne}, \citenamefont {Baba},
  \citenamefont {Clericò}, \citenamefont {Vila}, \citenamefont {Vaquero},
  \citenamefont {Delgado-Notario}, \citenamefont {Caridad}, \citenamefont
  {Watanabe}, \citenamefont {Taniguchi}, \citenamefont {Molina}, \citenamefont
  {Domínguez-Adame}, \citenamefont {Roche}, \citenamefont {Diez},
  \citenamefont {Rappoport},\ and\ \citenamefont
  {Amado}}]{Us-Experiment-salvadorsanchez2022generation}%
  \BibitemOpen
  \bibfield  {author} {\bibinfo {author} {\bibfnamefont {J.}~\bibnamefont
  {Salvador-Sánchez}}, \bibinfo {author} {\bibfnamefont {L.~M.}\ \bibnamefont
  {Canonico}}, \bibinfo {author} {\bibfnamefont {A.}~\bibnamefont
  {Pérez-Rodríguez}}, \bibinfo {author} {\bibfnamefont {T.~P.}\ \bibnamefont
  {Cysne}}, \bibinfo {author} {\bibfnamefont {Y.}~\bibnamefont {Baba}},
  \bibinfo {author} {\bibfnamefont {V.}~\bibnamefont {Clericò}}, \bibinfo
  {author} {\bibfnamefont {M.}~\bibnamefont {Vila}}, \bibinfo {author}
  {\bibfnamefont {D.}~\bibnamefont {Vaquero}}, \bibinfo {author} {\bibfnamefont
  {J.~A.}\ \bibnamefont {Delgado-Notario}}, \bibinfo {author} {\bibfnamefont
  {J.~M.}\ \bibnamefont {Caridad}}, \bibinfo {author} {\bibfnamefont
  {K.}~\bibnamefont {Watanabe}}, \bibinfo {author} {\bibfnamefont
  {T.}~\bibnamefont {Taniguchi}}, \bibinfo {author} {\bibfnamefont {R.~A.}\
  \bibnamefont {Molina}}, \bibinfo {author} {\bibfnamefont {F.}~\bibnamefont
  {Domínguez-Adame}}, \bibinfo {author} {\bibfnamefont {S.}~\bibnamefont
  {Roche}}, \bibinfo {author} {\bibfnamefont {E.}~\bibnamefont {Diez}},
  \bibinfo {author} {\bibfnamefont {T.~G.}\ \bibnamefont {Rappoport}},\ and\
  \bibinfo {author} {\bibfnamefont {M.}~\bibnamefont {Amado}},\ }\href@noop {}
  {\bibinfo {title} {Generation and control of non-local chiral currents in
  graphene superlattices by orbital hall effect}} (\bibinfo {year} {2022}),\
  \Eprint {https://arxiv.org/abs/2206.04565} {arXiv:2206.04565
  [cond-mat.mes-hall]} \BibitemShut {NoStop}%
\bibitem [{\citenamefont {Kohn}(1959)}]{OMM-Kohn-PhysRev.115.1460}%
  \BibitemOpen
  \bibfield  {author} {\bibinfo {author} {\bibfnamefont {W.}~\bibnamefont
  {Kohn}},\ }\bibfield  {title} {\bibinfo {title} {Theory of bloch electrons in
  a magnetic field: The effective hamiltonian},\ }\href
  {https://doi.org/10.1103/PhysRev.115.1460} {\bibfield  {journal} {\bibinfo
  {journal} {Phys. Rev.}\ }\textbf {\bibinfo {volume} {115}},\ \bibinfo {pages}
  {1460} (\bibinfo {year} {1959})}\BibitemShut {NoStop}%
\bibitem [{\citenamefont {Fa{\'{\i}}lde}\ and\ \citenamefont
  {Baldomir}(2021)}]{magField-Falde-NJPhys-2021}%
  \BibitemOpen
  \bibfield  {author} {\bibinfo {author} {\bibfnamefont {D.}~\bibnamefont
  {Fa{\'{\i}}lde}}\ and\ \bibinfo {author} {\bibfnamefont {D.}~\bibnamefont
  {Baldomir}},\ }\bibfield  {title} {\bibinfo {title} {Orbital dynamics in 2d
  topological and chern insulators},\ }\href
  {https://doi.org/10.1088/1367-2630/ac29fc} {\bibfield  {journal} {\bibinfo
  {journal} {New Journal of Physics}\ }\textbf {\bibinfo {volume} {23}},\
  \bibinfo {pages} {113002} (\bibinfo {year} {2021})}\BibitemShut {NoStop}%
\bibitem [{\citenamefont {Cai}\ \emph {et~al.}(2013)\citenamefont {Cai},
  \citenamefont {Yang}, \citenamefont {Li}, \citenamefont {Zhang},
  \citenamefont {Shi}, \citenamefont {Yao},\ and\ \citenamefont
  {Niu}}]{QNiu-PhysRevB.88.115140}%
  \BibitemOpen
  \bibfield  {author} {\bibinfo {author} {\bibfnamefont {T.}~\bibnamefont
  {Cai}}, \bibinfo {author} {\bibfnamefont {S.~A.}\ \bibnamefont {Yang}},
  \bibinfo {author} {\bibfnamefont {X.}~\bibnamefont {Li}}, \bibinfo {author}
  {\bibfnamefont {F.}~\bibnamefont {Zhang}}, \bibinfo {author} {\bibfnamefont
  {J.}~\bibnamefont {Shi}}, \bibinfo {author} {\bibfnamefont {W.}~\bibnamefont
  {Yao}},\ and\ \bibinfo {author} {\bibfnamefont {Q.}~\bibnamefont {Niu}},\
  }\bibfield  {title} {\bibinfo {title} {Magnetic control of the valley degree
  of freedom of massive dirac fermions with application to transition metal
  dichalcogenides},\ }\href {https://doi.org/10.1103/PhysRevB.88.115140}
  {\bibfield  {journal} {\bibinfo  {journal} {Phys. Rev. B}\ }\textbf {\bibinfo
  {volume} {88}},\ \bibinfo {pages} {115140} (\bibinfo {year}
  {2013})}\BibitemShut {NoStop}%
\bibitem [{\citenamefont {Xiao}\ \emph {et~al.}(2005)\citenamefont {Xiao},
  \citenamefont {Shi},\ and\ \citenamefont
  {Niu}}]{ModTheo-PhysRevLett.95.137204}%
  \BibitemOpen
  \bibfield  {author} {\bibinfo {author} {\bibfnamefont {D.}~\bibnamefont
  {Xiao}}, \bibinfo {author} {\bibfnamefont {J.}~\bibnamefont {Shi}},\ and\
  \bibinfo {author} {\bibfnamefont {Q.}~\bibnamefont {Niu}},\ }\bibfield
  {title} {\bibinfo {title} {Berry phase correction to electron density of
  states in solids},\ }\href {https://doi.org/10.1103/PhysRevLett.95.137204}
  {\bibfield  {journal} {\bibinfo  {journal} {Phys. Rev. Lett.}\ }\textbf
  {\bibinfo {volume} {95}},\ \bibinfo {pages} {137204} (\bibinfo {year}
  {2005})}\BibitemShut {NoStop}%
\bibitem [{\citenamefont {Kumar}\ and\ \citenamefont
  {Kumar}(2023)}]{IOHE-THzEmission-Kumar2023-NatComm}%
  \BibitemOpen
  \bibfield  {author} {\bibinfo {author} {\bibfnamefont {S.}~\bibnamefont
  {Kumar}}\ and\ \bibinfo {author} {\bibfnamefont {S.}~\bibnamefont {Kumar}},\
  }\bibfield  {title} {\bibinfo {title} {Ultrafast thz probing of nonlocal
  orbital current in transverse multilayer metallic heterostructures},\
  }\bibfield  {journal} {\bibinfo  {journal} {Nature Communications}\ }\textbf
  {\bibinfo {volume} {14}},\ \href {https://doi.org/10.1038/s41467-023-43956-y}
  {10.1038/s41467-023-43956-y} (\bibinfo {year} {2023})\BibitemShut {NoStop}%
\bibitem [{\citenamefont {Kampfrath}\ \emph {et~al.}(2013)\citenamefont
  {Kampfrath}, \citenamefont {Battiato}, \citenamefont {Maldonado},
  \citenamefont {Eilers}, \citenamefont {N\"{o}tzold}, \citenamefont
  {M\"{a}hrlein}, \citenamefont {Zbarsky}, \citenamefont {Freimuth},
  \citenamefont {Mokrousov}, \citenamefont {Bl\"{u}gel}, \citenamefont {Wolf},
  \citenamefont {Radu}, \citenamefont {Oppeneer},\ and\ \citenamefont
  {M\"{u}nzenberg}}]{Oppeneer-10.1038/nnano.2013.43}%
  \BibitemOpen
  \bibfield  {author} {\bibinfo {author} {\bibfnamefont {T.}~\bibnamefont
  {Kampfrath}}, \bibinfo {author} {\bibfnamefont {M.}~\bibnamefont {Battiato}},
  \bibinfo {author} {\bibfnamefont {P.}~\bibnamefont {Maldonado}}, \bibinfo
  {author} {\bibfnamefont {G.}~\bibnamefont {Eilers}}, \bibinfo {author}
  {\bibfnamefont {J.}~\bibnamefont {N\"{o}tzold}}, \bibinfo {author}
  {\bibfnamefont {S.}~\bibnamefont {M\"{a}hrlein}}, \bibinfo {author}
  {\bibfnamefont {V.}~\bibnamefont {Zbarsky}}, \bibinfo {author} {\bibfnamefont
  {F.}~\bibnamefont {Freimuth}}, \bibinfo {author} {\bibfnamefont
  {Y.}~\bibnamefont {Mokrousov}}, \bibinfo {author} {\bibfnamefont
  {S.}~\bibnamefont {Bl\"{u}gel}}, \bibinfo {author} {\bibfnamefont
  {M.}~\bibnamefont {Wolf}}, \bibinfo {author} {\bibfnamefont {I.}~\bibnamefont
  {Radu}}, \bibinfo {author} {\bibfnamefont {P.~M.}\ \bibnamefont {Oppeneer}},\
  and\ \bibinfo {author} {\bibfnamefont {M.}~\bibnamefont {M\"{u}nzenberg}},\
  }\bibfield  {title} {\bibinfo {title} {Terahertz spin current pulses
  controlled by magnetic heterostructures},\ }\href
  {https://doi.org/10.1038/nnano.2013.43} {\bibfield  {journal} {\bibinfo
  {journal} {Nature Nanotechnology}\ }\textbf {\bibinfo {volume} {8}},\
  \bibinfo {pages} {256–260} (\bibinfo {year} {2013})}\BibitemShut {NoStop}%
\bibitem [{\citenamefont {Guimar\~aes}\ \emph {et~al.}(2015)\citenamefont
  {Guimar\~aes}, \citenamefont {Lounis}, \citenamefont {Costa},\ and\
  \citenamefont {Muniz}}]{Bechara-PhysRevB.92.220410}%
  \BibitemOpen
  \bibfield  {author} {\bibinfo {author} {\bibfnamefont {F.~S.~M.}\
  \bibnamefont {Guimar\~aes}}, \bibinfo {author} {\bibfnamefont
  {S.}~\bibnamefont {Lounis}}, \bibinfo {author} {\bibfnamefont {A.~T.}\
  \bibnamefont {Costa}},\ and\ \bibinfo {author} {\bibfnamefont {R.~B.}\
  \bibnamefont {Muniz}},\ }\bibfield  {title} {\bibinfo {title} {Dynamical
  current-induced ferromagnetic and antiferromagnetic resonances},\ }\href
  {https://doi.org/10.1103/PhysRevB.92.220410} {\bibfield  {journal} {\bibinfo
  {journal} {Phys. Rev. B}\ }\textbf {\bibinfo {volume} {92}},\ \bibinfo
  {pages} {220410} (\bibinfo {year} {2015})}\BibitemShut {NoStop}%
\bibitem [{\citenamefont {Reiss}\ and\ \citenamefont
  {Brouwer}(2022)}]{Brouwer-PhysRevB.106.144423}%
  \BibitemOpen
  \bibfield  {author} {\bibinfo {author} {\bibfnamefont {D.~A.}\ \bibnamefont
  {Reiss}}\ and\ \bibinfo {author} {\bibfnamefont {P.~W.}\ \bibnamefont
  {Brouwer}},\ }\bibfield  {title} {\bibinfo {title} {Finite-frequency spin
  conductance of the interface between a ferro- or ferrimagnetic insulator and
  a normal metal},\ }\href {https://doi.org/10.1103/PhysRevB.106.144423}
  {\bibfield  {journal} {\bibinfo  {journal} {Phys. Rev. B}\ }\textbf {\bibinfo
  {volume} {106}},\ \bibinfo {pages} {144423} (\bibinfo {year}
  {2022})}\BibitemShut {NoStop}%
\bibitem [{\citenamefont {Reiss}\ \emph {et~al.}(2021)\citenamefont {Reiss},
  \citenamefont {Kampfrath},\ and\ \citenamefont
  {Brouwer}}]{Brouwer-PhysRevB.104.024415}%
  \BibitemOpen
  \bibfield  {author} {\bibinfo {author} {\bibfnamefont {D.~A.}\ \bibnamefont
  {Reiss}}, \bibinfo {author} {\bibfnamefont {T.}~\bibnamefont {Kampfrath}},\
  and\ \bibinfo {author} {\bibfnamefont {P.~W.}\ \bibnamefont {Brouwer}},\
  }\bibfield  {title} {\bibinfo {title} {Theory of spin-hall magnetoresistance
  in the ac terahertz regime},\ }\href
  {https://doi.org/10.1103/PhysRevB.104.024415} {\bibfield  {journal} {\bibinfo
   {journal} {Phys. Rev. B}\ }\textbf {\bibinfo {volume} {104}},\ \bibinfo
  {pages} {024415} (\bibinfo {year} {2021})}\BibitemShut {NoStop}%
\bibitem [{\citenamefont {Weiler}\ \emph
  {et~al.}(2014{\natexlab{b}})\citenamefont {Weiler}, \citenamefont {Shaw},
  \citenamefont {Nembach},\ and\ \citenamefont
  {Silva}}]{Silva-PhysRevLett.113.157204}%
  \BibitemOpen
  \bibfield  {author} {\bibinfo {author} {\bibfnamefont {M.}~\bibnamefont
  {Weiler}}, \bibinfo {author} {\bibfnamefont {J.~M.}\ \bibnamefont {Shaw}},
  \bibinfo {author} {\bibfnamefont {H.~T.}\ \bibnamefont {Nembach}},\ and\
  \bibinfo {author} {\bibfnamefont {T.~J.}\ \bibnamefont {Silva}},\ }\bibfield
  {title} {\bibinfo {title} {Phase-sensitive detection of spin pumping via the
  ac inverse spin hall effect},\ }\href
  {https://doi.org/10.1103/PhysRevLett.113.157204} {\bibfield  {journal}
  {\bibinfo  {journal} {Phys. Rev. Lett.}\ }\textbf {\bibinfo {volume} {113}},\
  \bibinfo {pages} {157204} (\bibinfo {year} {2014}{\natexlab{b}})}\BibitemShut
  {NoStop}%
\bibitem [{\citenamefont {Walowski}\ and\ \citenamefont
  {Münzenberg}(2016)}]{Walowski-10.1063_1.4958846}%
  \BibitemOpen
  \bibfield  {author} {\bibinfo {author} {\bibfnamefont {J.}~\bibnamefont
  {Walowski}}\ and\ \bibinfo {author} {\bibfnamefont {M.}~\bibnamefont
  {Münzenberg}},\ }\bibfield  {title} {\bibinfo {title} {Perspective:
  Ultrafast magnetism and thz spintronics},\ }\href
  {https://doi.org/10.1063/1.4958846} {\bibfield  {journal} {\bibinfo
  {journal} {Journal of Applied Physics}\ }\textbf {\bibinfo {volume} {120}},\
  \bibinfo {pages} {140901} (\bibinfo {year} {2016})}\BibitemShut {NoStop}%
\bibitem [{\citenamefont {Chang}\ and\ \citenamefont
  {Niu}(1996)}]{OMM-QNiu-PhysRevB.53.7010}%
  \BibitemOpen
  \bibfield  {author} {\bibinfo {author} {\bibfnamefont {M.-C.}\ \bibnamefont
  {Chang}}\ and\ \bibinfo {author} {\bibfnamefont {Q.}~\bibnamefont {Niu}},\
  }\bibfield  {title} {\bibinfo {title} {Berry phase, hyperorbits, and the
  hofstadter spectrum: Semiclassical dynamics in magnetic bloch bands},\ }\href
  {https://doi.org/10.1103/PhysRevB.53.7010} {\bibfield  {journal} {\bibinfo
  {journal} {Phys. Rev. B}\ }\textbf {\bibinfo {volume} {53}},\ \bibinfo
  {pages} {7010} (\bibinfo {year} {1996})}\BibitemShut {NoStop}%
\bibitem [{\citenamefont {Culcer}\ \emph {et~al.}(2005)\citenamefont {Culcer},
  \citenamefont {Yao},\ and\ \citenamefont {Niu}}]{Culcer-PhysRevB.72.085110}%
  \BibitemOpen
  \bibfield  {author} {\bibinfo {author} {\bibfnamefont {D.}~\bibnamefont
  {Culcer}}, \bibinfo {author} {\bibfnamefont {Y.}~\bibnamefont {Yao}},\ and\
  \bibinfo {author} {\bibfnamefont {Q.}~\bibnamefont {Niu}},\ }\bibfield
  {title} {\bibinfo {title} {Coherent wave-packet evolution in coupled bands},\
  }\href {https://doi.org/10.1103/PhysRevB.72.085110} {\bibfield  {journal}
  {\bibinfo  {journal} {Phys. Rev. B}\ }\textbf {\bibinfo {volume} {72}},\
  \bibinfo {pages} {085110} (\bibinfo {year} {2005})}\BibitemShut {NoStop}%
\end{thebibliography}

%

\end{document}